\documentclass[aip, cha, 11pt]{revtex4}
\pdfoutput=1
\usepackage{graphicx}

\usepackage{amsmath}
\usepackage{amssymb}
\usepackage{bm}

\newcommand{\la}{\left\langle}
\newcommand{\ra}{\right\rangle}

\newcommand{\abs}[1]{{|#1|}}

\begin{document}

\title{
  Fast optimal entrainment of limit-cycle oscillators by strong periodic inputs via phase-amplitude reduction and Floquet theory
}

\author{Shohei Takata}
\thanks{E-mail: takata.s.ae@m.titech.ac.jp}
\affiliation{Department of Systems and Control Engineering, Tokyo Institute of Technology, Tokyo 152-8552, Japan}

\author{Yuzuru Kato}
\thanks{Corresponding author. E-mail: kato.y.bg@m.titech.ac.jp}
\affiliation{Department of Systems and Control Engineering, Tokyo Institute of Technology, Tokyo 152-8552, Japan}

\author{Hiroya Nakao}
\thanks{E-mail: nakao@sc.e.titech.ac.jp}
\affiliation{Department of Systems and Control Engineering, Tokyo Institute of Technology, Tokyo 152-8552, Japan}

\date{\today}

\begin{abstract}
Optimal entrainment of limit-cycle oscillators by strong periodic inputs is studied on the basis of the phase-amplitude reduction and Floquet theory.
Two methods for deriving the input waveforms that keep the system state close to the original limit cycle are proposed, which enable the use of strong inputs for entrainment.
The first amplitude-feedback method uses feedback control to suppress deviations of the system state from the limit cycle, while the second amplitude-penalty method seeks an input waveform that does not excite large deviations from the limit cycle in the feedforward framework.
Optimal entrainment of the van der Pol and Willamowski-R\"ossler oscillators with real or complex Floquet exponents are analyzed as examples. It is demonstrated that the proposed methods can achieve considerably faster entrainment and provide wider entrainment ranges than the conventional method that relies only on phase reduction.
\end{abstract}

\maketitle

%%%%%%%%%%%%%%%%%%%%%%
%%% Lead paragraph %%%
%%%%%%%%%%%%%%%%%%%%%%

\hspace{1cm}

{\bf Entrainment of self-sustained oscillators by periodic inputs is widely observed in the real world, including the entrainment of circadian rhythms to sunlight and injection locking of electrical oscillators to clock signals. Optimization of input waveforms for stable entrainment has been considered by using phase reduction, which neglects amplitude deviations of the oscillator state from the original orbit. However, such methods do not perform well for strong inputs because the phase-only approximation breaks down. In this study, using phase-amplitude reduction, we propose two methods for obtaining input waveforms that can suppress amplitude deviations. We demonstrate that both methods enable us to achieve faster entrainment by applying stronger periodic inputs.}

\hspace{1cm}

%%%%%%%%%%%%%%%%%
%%% Section 1 %%%
%%%%%%%%%%%%%%%%%

\section{Introduction}

Synchronization or entrainment is a phenomenon in which rhythms of self-sustained oscillators adjust with each other or adjust to periodic external inputs. It is observed in a variety of real-world systems including metronomes~\cite{pantaleone2002synchronization}, Belousov-Zhabotinsky chemical reaction~\cite{winfree1972spiral}, flashing fireflies~\cite{buck1968mechanism, buck1976synchronous, ermentrout1984beyond}, circadian rhythms~\cite{goldbeter1995model, leloup1999limit}, and many others~\cite{winfree2001geometry, kuramoto1984chemical, ermentrout2010mathematical, pikovsky2001synchronization, glass1988clocks, strogatz1994nonlinear}.
Entrainment of nonlinear limit-cycle oscillators by external periodic inputs, also known as injection locking, finds many technological applications such as the frequency tuning in millimeter-wave oscillators~\cite{kawasaki2010millimeter, daryoush1990optical}, frequency stabilization of class-E electrical oscillators~\cite{nagashima2014locking}, suppression of pulsus alternans in the heart~\cite{wilson2017spatiotemporal, monga2019optimal}, and adjustment of circadian rhythms~\cite{leloup1999limit, stone2019application}.

When the input given to the limit-cycle oscillator is sufficiently weak, the phase reduction theory can be used to analyze the oscillator dynamics~\cite{winfree2001geometry,kuramoto1984chemical,ermentrout2010mathematical,nakao2016phase,monga2019phase,kuramoto2019concept}, which allows us to describe multidimensional nonlinear dynamics of the oscillator by an approximate one-dimensional phase equation. It has been extensively used in the analysis of coupled oscillators~\cite{pikovsky2001synchronization}, in particular, in explaining their collective synchronized dynamics~\cite{kuramoto1984chemical, acebron2005kuramoto, strogatz2000kuramoto}.
Recently, the phase reduction theory has been formulated also for non-conventional physical systems, such as piecewise-smooth oscillators~\cite{shirasaka2017phase2}, oscillators with time delay~\cite{kotani2012adjoint}, rhythmic spatiotemporal patterns~\cite{kawamura2013collective,nakao2014phase}, and quantum limit-cycle oscillators~\cite{kato2019semiclassical}.

The phase equation can also be used to formulate optimization and control problems for limit-cycle oscillators~\cite{monga2019phase}, for example, minimizing the control power~\cite{moehlis2006optimal, dasanayake2011optimal, zlotnik2012optimal, li2013control}, maximizing the locking range~\cite{harada2010optimal, tanaka2014optimal, tanaka2015optimal} and linear stability~\cite{zlotnik2013optimal} in the entrainment, maximizing the linear stability of mutual synchronization between coupled oscillators~\cite{shirasaka2017optimizing, watanabe2019optimization}, maximizing the phase coherence of noisy oscillators~\cite{pikovsky2015maximizing}, performing phase-selective entrainment of oscillators~\cite{zlotnik2016phase}, and controlling the phase distributions in oscillator populations~\cite{monga2018synchronizing, kuritz2019ensemble, monga2019phase2, kato2021optimization}. 

Optimization methods based on phase reduction have also been studied in non-conventional oscillatory systems such as mutual synchronization between rhythmic spatiotemporal patterns~\cite{kawamura2017optimizing} and collectively oscillating networks~\cite{yamaguchi2021network}, and entrainment of a quantum limit-cycle oscillator in the semiclassical regime~\cite{kato2020semiclassical}.
However, these methods do not perform well when the driving input given to the oscillator is not sufficiently weak because the amplitude deviations of the system state from the unperturbed limit cycle become non-negligible and the phase-only description of the system breaks down; this is a fundamental limitation hampering practical applications of all methods that rely only on phase reduction.

Recent developments in the Koopman operator theory~\cite{mezic2005spectral, mezic2013analysis} have clarified that the deviations of the system state from the limit cycle can be characterized naturally by the amplitude variables, which are Koopman eigenfunctions associated with non-zero Floquet exponents of the system \cite{mauroy2013isostables,mauroy2014global}.
Phase-amplitude reduction theories that generalize the conventional phase-only reduction theory by including the amplitude variables have also been formulated~\cite{wilson2016isostable, mauroy2016global, mauroy2018global, shirasaka2017phase, shirasaka2020phase, monga2019phase, kotani2020nonlinear, nakao2021phaseamplitude}.
By using the resulting phase-amplitude equations, optimization methods have been proposed~\cite{shirasaka2020phase, monga2019optimal, wilson2018greater, wilson2021optimal}, for example, for minimizing 
the control power applied to the oscillator with slow amplitude relaxation.

In this study, on the basis of the phase-amplitude reduction and Floquet theory, 
we propose two methods that allow us to apply strong inputs to the oscillator and realize fast entrainment:
an {\it amplitude-feedback method}, which applies feedback control to suppress amplitude deviations in addition to the optimal input for phase locking, and an {\it amplitude-penalty method}, which is in the feedforward framework and uses the optimal input that 
is designed not to excite large amplitude deviations.
Using the van der Pol and Willamowski-R\"ossler  oscillators  as examples, we demonstrate that the proposed methods can significantly improve the convergence time to the entrained state and enlarge the entrainment range.

This paper is organized as follows; in Sec.~II, the phase-amplitude reduction and Floquet theory are briefly explained; in Sec.~III, the conventional phase-only optimization method for stable entrainment is reviewed and its failure for strong inputs is demonstrated; in Sec.~IV, the amplitude-feedback and amplitude-penalty methods are proposed on the basis of phase-amplitude reduction; in Sec.~V, the validity of the proposed methods are demonstrated by using the van der Pol and Willamowski-R\"ossler oscillators; in Sec.~VI, conclusions and discussion are given, and in the Appendix, additional discussions on the optimal tangential input and details of the numerical methods for calculating the Floquet eigenvectors are given.

%%%%%%%%%%%%%%%%%
%%% Section 2 %%%
%%%%%%%%%%%%%%%%%

\section{Phase-amplitude reduction and Floquet theory}

In this section, we briefly review the phase-amplitude reduction theory for limit-cycle oscillators and its relation to the Floquet theory.

%%%%%%%%%%%%%%%%%%%%%%%%%%%%%%%%%%%%%%%%%%%%%%%%
\subsection{Phase-amplitude reduction}
%%%%%%%%%%%%%%%%%%%%%%%%%%%%%%%%%%%%%%%%%%%%%%%%

We begin with a brief introduction of the phase-amplitude reduction \cite{wilson2016isostable, mauroy2018global, shirasaka2017phase, shirasaka2020phase}. We consider a limit-cycle oscillator described by 
\begin{align}
  \label{eq:x}
  \dot{\bm X}(t) = \bm{F}({\bm X}(t)),
\end{align}
where ${\bm X}(t) \in \mathbb{R}^{N}$ is the system state at time $t$, 
the dot $(\dot{})$ denotes the time derivative, and
$\bm{F}({\bm X})$ $\in \mathbb{R}^{N}$ describes the system dynamics.
It is assumed that Eq.~(\ref{eq:x}) has 
an exponentially stable limit-cycle solution ${\bm X}_{0}(t) = {\bm X}_{0}(t+T)$ with period $T$
$=2\pi/\omega$, where $\omega$ is the natural frequency. We denote this limit cycle as $\chi$.

The linear stability of $\chi$ is characterized by the Floquet exponents $\lambda_i$ ($i=0, 1, ..., N-1$). One of the exponents vanishes, $\lambda_0 = 0$, which is associated with the phase direction tangent to $\chi$, and the other exponents $\lambda_1, ..., \lambda_{N-1}$ associated with the amplitude deviations from $\chi$ possess negative real parts and are in general complex.
These Floquet exponents are sorted in decreasing order of their real parts, i.e., $\lambda_0 = 0 > \mbox{Re}({\lambda_1}) \geq  \mbox{Re}({\lambda_2}) \geq \ldots  \geq \mbox{Re}({\lambda_{N-1}})$.

For this system, we can define a phase function $\Theta({\bm X}) : \mathbb{R}^{N} \to [0, 2\pi)$ and amplitude functions $R_i({\bm X}) : \mathbb{R}^{N} \to {\mathbb C}~(i = 1, \ldots, N-1)$ in the basin of $\chi$ satisfying
\begin{align}
  \label{eq:def_phi_r}
  \dot{\Theta}({\bm X}) &= \la \nabla \Theta({\bm X}), \bm{F}({\bm X})  \ra  = \omega,
  \cr
  \dot{R}_i({\bm X}) &= \la   \nabla R_i({\bm X}),  \bm{F}({\bm X}) \ra = \lambda_i R_i({\bm X}),
\end{align}
where the inner product is defined by $ \la \bm{a}, \bm{b} \ra = \sum_{j = 1}^{N} a^*_{j} b_{j}$ ($*$ denotes complex conjugate).
It is noted that the phase constantly increases with the frequency $\omega$ and the amplitudes decay linearly with the rate (possibly complex) $\lambda_i$, namely, the phase and amplitude functions provide a global linearization of the original nonlinear system in the basin of $\chi$~\cite{mauroy2013isostables}.

Using the phase and amplitude functions $\Theta$ and $R_i$, we introduce the phase and amplitude variables of the system state ${\bm X}$ as $\theta = \Theta({\bm X})$ and $r_i = R_i({\bm X})$ $(i=1, ..., N-1$). When the state ${\bm X}(t)$ is on $\chi$, it can be represented as a function of the phase as ${\bm X}(t) = {\bm X}_0(\theta/\omega) =: {\bm \chi}(\theta) $, where the state ${\bm X}_0(0)$ at $t=0$ is chosen as the origin $\theta=0$ of the phase without loss of generality, and the symbol ${\bm \chi}(\theta)$ is introduced to represent the state on the limit cycle $\chi$ as a function of the phase $\theta$. Each amplitude $R_i({\bm X})$ measures the deviation of the system state ${\bm X}$ from $\chi$ and vanishes when ${\bm X}$ is on $\chi$, i.e., $R_i({\bm \chi}(\theta))=0$.

The phase defined above is called the {\it asymptotic phase}~\cite{winfree2001geometry, kuramoto1984chemical, pikovsky2001synchronization, ermentrout2010mathematical,nakao2016phase,kuramoto2019concept} and characterizes the oscillator dynamics along $\chi$. The level sets of this phase function is called {\it isochrons}. Similarly, the level sets of the (absolute values of the) amplitudes characterizing deviations of the system state from $\chi$ are called {\it isostables}~\cite{mauroy2013isostables}.

We now assume that the oscillator is perturbed by an external input as
\begin{align}
  \label{eq:xperturb}
%  \dot{\bm X}(t) = \bm{F}({\bm X}(t)) + \bm{p}({\bm X}(t), t),
   \dot{\bm X}(t) = \bm{F}({\bm X}(t)) + \bm{p}(t),
\end{align}
where ${\bm p}(t) \in \mathbb{R}^{N}$ represents the input.
By using the chain rule of differentiation, we can derive the equations for the phase and amplitudes variables as
\begin{align}
  \label{eq:phi_r}
  \dot{\theta}(t) &= \omega +  \la \nabla \Theta({\bm X}(t)), \bm{p}(t) \ra,
  \cr
  \dot{r}_i(t) &= \lambda_i r_i(t) + \la \nabla R_i({\bm X}(t)),  {\bm p}(t) \ra.
\end{align}
We are interested in the phase $\theta$ and only a small number of slowly decaying amplitudes $r_1, ..., r_M$ 
($M < N-1$) to reduce the dimensionality of the system.

In the conventional phase(-amplitude) reduction for a weakly perturbed limit-cycle oscillator~\cite{kuramoto1984chemical, nakao2016phase}, it is assumed that the perturbation ${\bm p}$ is sufficiently weak and of order $\mathcal{O}(\epsilon)$ where $0 \leq \epsilon \ll 1$. 
Then, neglecting the terms of order $\mathcal{O}(\epsilon^2)$,
we can obtain the approximate phase-amplitude equations at the lowest order in a closed form (see e.g. Refs.~\cite{wilson2016isostable,shirasaka2017phase,kuramoto2019concept}),
\begin{align}
  \dot{\theta} &= \omega +  \la \bm{Z}(\theta),  \bm{p}(t) \ra,
  \cr
  \dot{r}_i &= \lambda_i r_i+   \la \bm{I}_i(\theta),  \bm{p}(t) \ra,
  \label{eq:phi-r-lowest}
\end{align}
where the gradients $\nabla \Theta({\bm X})$ and $\nabla R_i({\bm X})$ are approximated by the phase sensitivity function (PSF) $\bm{Z}(\theta) = \nabla\Theta|_{ {\bm X} = {\bm \chi}(\theta) }$ and amplitude (isostable) sensitivity functions (ISFs) $\bm{I}_i(\theta) = \nabla R_i |_{ {\bm X} = {\bm \chi}(\theta) }$ evaluated at ${\bm X} = {\bm \chi}(\theta)$ on the limit cycle $\chi$.

In this study, however, we do not assume that the perturbation ${\bm p}$ is  weak in deriving the phase-amplitude equations. Instead, we assume that each amplitude $r_i$ is kept to be of order $\mathcal{O}({\delta})$ with $0 \leq {\delta} \ll 1$ even under the effect of ${\bm p}$, namely, ${\bm p}$ is chosen such that the amplitude $r_i$ does not deviate largely from $0$ as we explain later
{(as such ${\bm p}$, we will use a periodic input with an additional feedback control, or a periodic input that is designed not to excite large amplitude deviations)}.
As shown in Appendix A,  we can then approximately derive the following phase-amplitude equations:
\begin{align}
  \label{eq:pr_phi_r}
  \dot{\theta} &= \omega +  \la \bm{Z}(\theta),  \bm{p}(  t) \ra + \mathcal{O}(\delta / \omega) ,
  \cr
  \dot{r}_i &= \lambda_i r_i+   \la \bm{I}_i(\theta),  \bm{p}( t) \ra +  \mathcal{O}(\delta / \omega),
\end{align}
Thus, even if ${\bm p}$ is not very small, we can neglect the terms of order $\mathcal{O}(\delta / \omega)$ when $\delta / \omega$ is small and obtain the lowest-order phase-amplitude equations of the form~(\ref{eq:phi-r-lowest}) also in this case.

It often happens that the negative real part of $\lambda_i$ decreases quickly with $i$ and only the first few amplitude variables $r_i$ for $i=1, ..., M$ decay slowly. 
In the simplest situation, we may need to consider only the slowest-decaying amplitude $r_1$ when $\lambda_1$ is real, or only a pair of slowest-decaying amplitudes $r_1$ and $r_2$ when $\lambda_1 = \lambda_2^*$ is complex, and reduce the dimensionality of the original $N$-dimensional system to $M+1 = 2$ or $3$.

From the phase and amplitudes $\theta$ and $r_i$, the original system state can be approximately reconstructed as ${\bm X} \approx {\bm \chi}(\theta)$ at the lowest order, or
\begin{align}
  {\bm X} \approx {\bm \chi}(\theta) + \sum_{i=1}^{M} r_i {\bm u}_i(\theta/\omega)
\end{align}
at the first order, where ${\bm u}_i(t)$ is the right Floquet eigenvector associated with $\lambda_i$ given in the next subsection.

%%%%%%%%%%%%%%%%%%%%%%%%%%%%%%%%%%%%%%%%%%%%%%%%
\subsection{Floquet theory}
%%%%%%%%%%%%%%%%%%%%%%%%%%%%%%%%%%%%%%%%%%%%%%%%

The phase-amplitude equations~(\ref{eq:phi-r-lowest}) are characterized by the natural frequency $\omega$, PSF $\bm{Z}(\theta)$, Floquet exponents $\lambda_i$, and ISFs $\bm{I}_i(\theta)$ for $i=1, ..., M$.
Though it is difficult to obtain the phase and amplitude functions $\Theta$ and $R_i$ analytically, the PSF and ISFs are given by the left Floquet eigenvectors of the limit cycle and can be evaluated more easily~\cite{ermentrout1996type, brown2004phase, winfree2001geometry, kuramoto1984chemical, pikovsky2001synchronization, nakao2016phase, ermentrout2010mathematical, shirasaka2017phase,kuramoto2019concept,wilson2016isostable}.

By linearizing the unperturbed system described by Eq.~(\ref{eq:x}) around the limit-cycle solution ${\bm X}_0(t)$ for a small variation ${\bm y}(t) = {\bm X}(t) - {\bm X}_0(t) \in \mathbb{R}^{N}$, we obtain a periodically driven linear system, $\dot{\bm y}(t) = {\bm J}({\bm X}_0(t)) {\bm y}(t)$, where ${\bm J} \in {\mathbb R}^{N \times N}$ is a $T$-periodic Jacobian matrix of ${\bm{F}}$ evaluated at ${\bm X} = {\bm X}_0(t)$. 
We denote the fundamental matrix  of this linearized equation as $\bm{\Psi}(t) \in {\mathbb R}^{N \times N}$, which satisfies 
$\dot{\bm{\Psi}}(t) = {\bm J}({\bm X}_{0}(t)) \bm{\Psi}(t) $
with an initial condition $\bm{\Psi}(0) = \bm{E}$, where $\bm{E} \in {\mathbb R}^{N \times N}$ is an identity matrix~\cite{jordan1999nonlinear,guckenheimer2013nonlinear}.

According to the Floquet theory, $\bm{\Psi}(t)$ can be expressed as
$\bm{\Psi}(t) = \bm{P}(t)\exp( \bm{\Lambda} t)$ where ${\bm P}$ is a $T$-periodic matrix satisfying $\bm{P}(t + T) = \bm{P}(t) \in {\mathbb R}^{N \times N}$ and ${\bm P}(0) = {\bm E}$, and $\bm{\Lambda} \in {\mathbb R}^{N \times N}$ is a constant matrix. It is noted that ${\bm \Lambda}$ depends on the initial condition ${\bm X}_0(0)$ at $t=0$ of the system. The matrix ${\bm M} = \exp( {\bm \Lambda} T )$ is called a monodromy matrix~\cite{jordan1999nonlinear,guckenheimer2013nonlinear}.

We consider the eigensystem $\{ \lambda_{i} \in {\mathbb C},\ \bm{u}_{i} \in {\mathbb C}^{N},\ \bm{v}_{i} \in {\mathbb C}^{N}\}_{i=0, 1, ..., N-1}$
of $\bm{\Lambda}$ satisfying $\bm{\Lambda} \bm{u}_{i} = \lambda_{i} \bm{u}_{i}$ and 
$ \bm{\Lambda}^{\dag} \bm{v}_{i} = \lambda^{*}_{i} \bm{v}_{i}$ for $i=0, 1, ..., N-1$, where ${\dag}$ represents Hermitian conjugate (transpose for real matrices).
The eigenvalue $\lambda_i$ gives the $i$th Floquet exponent (i.e., $e^{\lambda_i T}$ is the Floquet multiplier; the principal value is chosen when $\lambda_i$ is complex) and we call the associated right and left eigenvectors ${\bm u}_i$ and ${\bm v}_i$ the Floquet eigenvectors.
Among them, one of the Floquet exponent is $\lambda_0 = 0$ and the associated right Floquet eigenvector ${\bm u}_0$ is parallel to the tangent vector of $\chi$ at ${\bm X}_0(0)$~\cite{kuramoto1984chemical,kuramoto2019concept}. We choose this vector as ${\bm u}_0 = (1/\omega) d{\bm X}_0(t)/dt|_{t=0}$, where the normalization factor $(1/\omega)$ is introduced to be consistent with the convention of the phase reduction theory~\cite{kuramoto1984chemical,kuramoto2019concept}.
The eigenvectors are normalized to satisfy the bi-orthonormality relation $ \la \bm{v}_{i}, \bm{u}_{j} \ra = {\delta}_{ij}$ for $i, j = 0, 1, ..., N$. 

Using the right and left Floquet eigenvectors ${\bm u}_i$ and ${\bm v}_i$, we further define
\begin{align}
{\bm{u}}_{i}(t) = \bm{P}(t) \bm{u}_{i} \in {\mathbb C}^{N},
\quad
{\bm{v}}_{i}(t) =  (\bm{P}(t)^{-1})^{\dag}\bm{v}_{i} \in {\mathbb C}^{N}
\label{floqueteigenvectors}
\end{align}
for $0 \leq t < T$ along $\chi$, which we also call the Floquet eigenvectors. These eigenvectors are $T$-periodic, i.e., ${\bm u}_i(T) = {\bm u}_i(0) = {\bm u}_i$ and ${\bm v}_i(T) = {\bm v}_i(0) = {\bm v}_i$, and satisfy the bi-orthonormality relation
$\la {\bm{v}}_{i}(t),  {\bm{u}}_{j}(t) \ra= {\delta}_{i, j}$
$(i, j=0, 1, ..., N-1)$
for $0 \leq t < T$.
We can also confirm that these vectors are $T$-periodic solutions to the following set of linear and adjoint linear equations~\cite{ermentrout1996type, brown2004phase, shirasaka2017phase, kuramoto2019concept}:
\begin{align}
  \label{eq:adj_eq}
  \frac{d}{dt}
{\bm{u}}_{i} (t) &= \left[ {\bm J}({\bm X}_0(t)) - \lambda_{i} \right] {\bm{u}}_{i}(t),
  \cr
  \frac{d}{dt}
{\bm{v}}_{i}(t) &= - \left[ {\bm J}({\bm X}_0(t))^{\dag} -  \lambda_{i}^{*} \right] {\bm{v}}_{i}(t).
\end{align}
In particular, the right Floquet eigenvector associated with $\lambda_0 = 0$ is given by ${\bm u}_0(t) = (1/\omega) d{\bm X}_0(t) / dt = (1/\omega) {\bm F}({\bm X}_0(t))$ and satisfies the above equation.

The PSF and ISFs, which are the gradients of the phase and amplitude functions $\Theta({\bm X})$ and $R_i({\bm X})$ evaluated at ${\bm X} = {\bm \chi}(\theta)$ with $\theta = \omega t$ on the limit cycle $\chi$, can be expressed by the left Floquet eigenvectors ${ {\bm v}}_i(t)$ ($i=0, 1, \ldots, N-1$) as
\begin{align}
	\label{eq:ZIvsV}
	{\bm Z}(\theta) &= \nabla\Theta |_{ {\bm X} = {\bm \chi}(\theta) } = {\bm{v}}_{0}(\theta / \omega),
\cr
	{\bm I}_i(\theta) &= \nabla R_{i} |_{ {\bm X} = {\bm \chi}(\theta) } = {\bm{v}}_{i}(\theta / \omega),
\end{align}
for $0 \leq \theta < 2\pi$. Indeed, assuming that the above relations hold, we can linearly approximate the phase and amplitude functions for ${\bm X}$ close to ${\bm \chi}(\theta)$ as
\begin{align}
  \Theta({\bm X}) &\approx \Theta({\bm \chi}(\theta)) +  \la \nabla\Theta|_{ {\bm X} = {\bm \chi}(\theta) },\ {\bm y}(t) \ra = \theta + \la {\bm{v}}_{0}(t),\ {\bm y}(t) \ra,
  \cr
  R_i({\bm X}) &\approx R_i({\bm \chi}(\theta)) +\la \nabla R_i|_{ {\bm X} = {\bm \chi}(\theta) },\   {\bm y}(t) \ra =  \la {\bm{v}}_{i}(t),\ {\bm y}(t) \ra,
\end{align}
for sufficiently small ${\bm y}(t) = {\bm X}(t) - {\bm \chi}(\theta(t))$.
We then have
\begin{align}
  \frac{d}{dt} \Theta({\bm X}) &\approx \frac{d\theta}{dt} + \la \frac{d {\bm{v}}_{0}( t)}{dt},\ {\bm y}(t) \ra + \la {\bm{v}}_{0}( t),\ \frac{d {\bm y}(t)}{dt} \ra
  \cr
  &= \omega + \la \frac{d {\bm{v}}_{0}( t)}{dt},\ {\bm y}(t) \ra + \la {\bm{v}}_{0}( t),\ {\bm J}({\bm X}_0(t)) {\bm y}(t) \ra
  \cr
  &= \omega + \la \frac{d {\bm{v}}_{0}( t)}{dt} + {\bm J}({\bm X}_0(t))^\dag {\bm v}_0(t),\ {\bm y}(t) \ra = \omega
\end{align}
and
\begin{align}
  \frac{d}{dt} R_i({\bm X}) &\approx \la \frac{d {\bm{v}}_{i}( t)}{dt},\ {\bm y}(t) \ra + \la {\bm{v}}_{i}( t),\ \frac{d {\bm y}(t)}{dt} \ra
  \cr
  &= \la \frac{d {\bm{v}}_{i}( t)}{dt},\ {\bm y}(t) \ra + \la {\bm{v}}_{i}( t),\ {\bm J}({\bm X}_0(t)) {\bm y}(t) \ra
  \cr
  &= \la \frac{d {\bm{v}}_{i}(t)}{dt} + {\bm J}({\bm X}_0(t))^\dag {\bm v}_i(t), {\bm y}(t) \ra = \la \lambda_i^* {\bm v}_i(t),\ {\bm y}(t) \ra = \lambda_i R_i({\bm X}),
\end{align}
hence $\Theta$ and $R_i$ satisfy their defining equations~(\ref{eq:def_phi_r}) within the linear approximation.
It is noted that, by the above definition of ${\bm u_0}(t)$ and ${\bm Z}(\theta)$, we have $d\Theta({\bm X}_0(t))/dt = \la \nabla \Theta({\bm X}_0(t)),\ d{\bm X}_0(t)/dt \ra = \la {\bm Z}(\theta), \omega {\bm u}_0(\theta) \ra = \omega \la {\bm v}_0(t), {\bm u}_0(t) \ra = \omega$, which is a standard convention in the phase reduction theory~\cite{kuramoto1984chemical,kuramoto2019concept}.
For more details, see e.g. Appendix A in Ref.~\cite{kuramoto2019concept}.

Thus, by numerically solving the linear and adjoint linear equations~(\ref{eq:adj_eq}) and obtaining ${\bm{u}}_i(t)$ and ${\bm{v}}_i(t)$, we can calculate the PSF ${\bm Z}(\theta)$ and ISF ${\bm I}_i(\theta)$ and use them to reduce the original $N$-dimensional dynamical system given by Eq.~(\ref{eq:x}) to the $(M+1)$-dimensional phase and amplitude equations~(\ref{eq:pr_phi_r}).
Details of the numerical schemes for calculating the Floquet eigenvectors are explained in Appendix B.

%%%%%%%%%%%%%%%%%
%%% Section 3 %%%
%%%%%%%%%%%%%%%%%

%\section{Optimal entrainment without amplitude suppression}
\section{Phase equation for a periodically driven oscillator}

In this section, we consider a limit-cycle oscillator driven by a periodic input. We first derive an averaged equation for the phase difference between the oscillator and the periodic input.
We then briefly review the optimization of the periodic input for linear stability within the phase-reduction framework by Zlotnik {\it et al.}~\cite{zlotnik2013optimal}. We also demonstrate that the method fails when the periodic input is too strong due to breakdown of the phase-only approximation, leading to discrepancy between the target and realized phase-locking points.

\subsection{Averaged phase equation}

We consider a limit-cycle oscillator subjected to a periodic input described by Eq.~(\ref{eq:xperturb}).
The perturbation is given in the form ${\bm{p}(t)} =  {\bm{q}(\Omega t)}$,  where ${\bm q} \in {\mathbb R}^N$ represents the waveform of the periodic input of frequency $\Omega$ and period $T_e = 2\pi/ \Omega$.
Following the previous studies for optimal entrainment~\cite{zlotnik2013optimal,harada2010optimal,tanaka2014optimal,tanaka2015optimal}, we first derive an approximate, autonomous equation for the phase difference $\phi(t) = \theta(t) - \Omega t$
by using the averaging method~\cite{kuramoto1984chemical,hoppensteadt1997weakly}, and then derive the optimal periodic waveform by using the averaged equation.

Unlike the conventional analysis~\cite{kuramoto1984chemical}, we do not assume the perturbation ${\bm q}$ to be small, but we assume that the functional form of ${\bm q}$ is chosen appropriately such that the amplitude deviations of the oscillator state from the limit cycle remains $\mathcal{O}({\delta})$ $(0 \leq \delta \ll 1)$ as explained in Sec.~II and Appendix A.
Here, to perform the averaging approximation, we additionally assume that ${\bm q}$ is of $\mathcal{O}(\omega \delta)$ and
that the natural frequency $\omega$ of the oscillator and the input frequency $\Omega$ are close to each other in the sense that the frequency mismatch $\Delta = \omega - \Omega$ is of order $\mathcal{O}(\omega \delta)$. We stress that ${\bm q}$ and $\Delta$ may not be small when $\omega$ is large.

The approximate equation for the oscillator phase $\theta(t) = \Theta({\bm X})$ is given from Eq.~(\ref{eq:pr_phi_r}) by
\begin{align}
  \frac{d}{dt} \theta(t) = \omega + \la \bm{Z}(\theta(t)),  {\bm{q}(\Omega t)} \ra,
  \label{isouq}
\end{align}
which is non-autonomous. To average this equation and derive an autonomous form, we introduce a rescaled time $t' = \Omega t$. Using $d/dt = \Omega d/dt'$, we obtain
\begin{align}
\Omega \frac{d}{dt'} \tilde\theta(t') = \omega + \la  {\bm Z}(\tilde\theta(t')),\ \tilde{{\bm q}}(t') \ra,
\end{align}
where $\tilde{\theta}(t') = \theta(t = t' / \Omega)$ and $\tilde{{\bm q}}(t') = {\bm q}(\Omega t = t')$.
The equation for the phase difference 
$\tilde{\phi}(t') = \phi(t = t' / \Omega) = \theta(t) - \Omega t = \tilde{\theta}(t') - t'$ 
is then given by
\begin{align}
\frac{d}{dt'} \tilde{\phi}(t') 
&= \frac{1}{\Omega} \left[ \Delta + \la {\bm Z}(\tilde\phi(t') + t'),\ \tilde{{\bm q}}(t') \ra \right].
\end{align}

Because we assumed  that ${\bm q}$ and $\Delta$ are of $\mathcal{O}(\omega \delta)$, the right-hand side of the above equation is of order $\mathcal{O}(\delta)$ and small, namely, $\tilde{\phi}$ is a slowly-varying quantity. We can thus average the right-hand side over one period of oscillation and derive an approximate equation for $\tilde{\phi}(t')$ as~\cite{kuramoto1984chemical,hoppensteadt1997weakly}
\begin{align}
	\frac{d}{dt'} \tilde{\phi}(t') 
	&= \frac{1}{\Omega} \left[ \frac{1}{T_e'} \int_0^{T_e'} \left\{ \Delta + \la {\bm Z}(\tilde\phi(t') + s'),
	\ \tilde{{\bm q}}(s') \ra \right\} ds' \right],
\end{align}
where $T_e' = \Omega T_e = 2 \pi$ is the input period measured in the timescale of $t'$ and a small error term of  
$\mathcal{O}(\delta^2)$ that
arises by the averaging was dropped~\cite{hoppensteadt1997weakly}.
Returning to the original timescale $t$, the approximate averaged equation for the phase difference $\phi(t) = \tilde\phi(t' = \Omega t)$ is given by
\begin{align}
\frac{d}{dt} \phi(t) = \Delta + \frac{1}{T_e} \int_0^{T_e} \la {\bm Z}(\phi + \Omega s),\ {{\bm q}}(\Omega s) \ra ds,
\end{align}
where we put $s' = \Omega s$ in the integral.

Thus, 
the approximate equation for the phase difference $\phi$ is given by
\begin{align}
  \dot{\phi} = \Delta  +  \Gamma(\phi),
\quad
\Gamma(\phi) =  \big[  \la \bm{Z}(\phi + \Omega s ),\ \bm{q}( \Omega s ) \ra \big]_s
  \label{phiT}
\end{align}
where we defined the phase coupling function $\Gamma(\phi)$ and $[ g(s) ]_{s} = \frac{1}{T_e} \int _0^{T_e} g(s) ds$ represents the time average of a given function $g \in {\mathbb R}$ over one input period $T_e$.
It is noted that, though our assumptions are different from the standard analysis for weakly perturbed oscillators, Eq.~(\ref{phiT}) for the phase difference takes the same form as the averaged phase equation in the standard analysis~\cite{kuramoto1984chemical}. Therefore, we can use the same optimization method as in Ref.~\cite{zlotnik2013optimal} to derive the optimal waveform for the periodic input ${\bm q}$.

\subsection{Optimal waveform for stable entrainment}

We seek the optimal input for which (i) a given $\phi^*$ $(0 \leq \phi^* < 2\pi)$ is a stable phase-locking point and (ii) the linear stability (negative of the stability exponent) at $\phi^*$,
\begin{align}
  -\Gamma'(\phi^*) =  - \big[  \la \bm{Z}'(\phi^* + \Omega t ),\  \bm{q}(\Omega t) \ra \big]_t,
\end{align}
is maximized under a constraint on the power of the periodic input ${\bm q}$.
This maximization problem is formulated as
\begin{align}
  \label{eq:opt_lins}
  &\max_{\bm{q}}\ -\Gamma'(\phi^*),
  \cr
  &\mbox{s.t.} \quad \Delta + \Gamma(\phi^*) = 0,\quad \big[ \| \bm{q} \|^2  \big]_t = P,
\end{align}
where the one-period average of the power $\| \bm{q} \|^2 = \la \bm{q}, \bm{q}\ra$ of the input is constrained at a constant $P > 0$, where  $\sqrt{P}$ is generally of order  $\mathcal{O}(\omega \delta)$.
These constraints can be incorporated into the Lagrange function
\begin{align}
  C(\bm{q}, \mu, \nu) = -\Gamma'(\phi^*) + \mu\{\Delta + \Gamma(\phi^*)\} + \nu(P - [\|\bm{q}\|^2]_t )
\end{align}
by introducing Lagrange multipliers $\mu$ and $\nu$.
From the conditions for the extremum, we obtain an Euler-Lagrange equation and two additional conditions as follows:
\begin{align}
  \frac{{\delta}{C}}{{\delta}{\bm{q}}} &= \frac{1}{T_e}\{-\bm{Z'}(\phi^*+\Omega t) + \mu \bm{Z}(\phi^*+ \Omega t) - 2\nu \bm{q}(\Omega t) \} = 0, \\
  \frac{\partial{C}}{\partial{\mu}} &= \Delta + [\la  \bm{Z}(\phi^*+\Omega t) ,  \bm{q}(\Omega t) \ra]_t = 0, \\
  \frac{\partial{C}}{\partial{\nu}} &= P - [ \|\bm{q}(\Omega t)\|^2 ]_t = 0,
\end{align}
where ${\delta}/{\delta}{{\bm q}}$ represents functional differentiation with respect to ${\bm q}$.
The optimal input waveform in the phase-reduction framework, denoted as $\bm{q}^{ls}(\Omega t)$, is thus obtained as
\begin{align}
  \bm{q}^{ls}(\Omega t) &= -\frac{1}{2\nu}\bm{Z'}(\phi^*+\Omega t) + \frac{\mu}{2\nu}\bm{Z}(\phi^*+\Omega t), 
  \label{eq:Zltonik_q} 
  \\
  \cr
  \mu &= - \frac{2\nu\Delta}{[ \| \bm{Z} \|^2 ]_t},
  \quad
  \nu = \frac{1}{2}\sqrt{\frac{  [\| \bm{Z'} \|^2]_t}{P- \frac{\Delta^2}{ [\| \bm{Z} \|^2]_t }}}.\label{eq:Zltonik_lambda}
\end{align}

\subsection{Breakdown of the phase-only approximation}

In the above derivation, we generally assumed that $\sqrt{P}$ of the periodic input is of order  $\mathcal{O}(\omega \delta)$. However, the phase equation is valid only when the amplitude deviations of the system state from the limit cycle remain small. Thus, if we simply apply the optimal waveform ${\bm q}^{ls}$ with non-small $P$ to the oscillator, the assumption of small amplitude deviations may be violated and the oscillator may not be entrained as desired because ${\bm q}^{ls}$ is not designed to suppress amplitude deviations.
Here, we illustrate the breakdown of the phase-only approximation for large $P$.

\begin{figure}[!t]
  \begin{center}
    \includegraphics[width=0.85\hsize,clip]{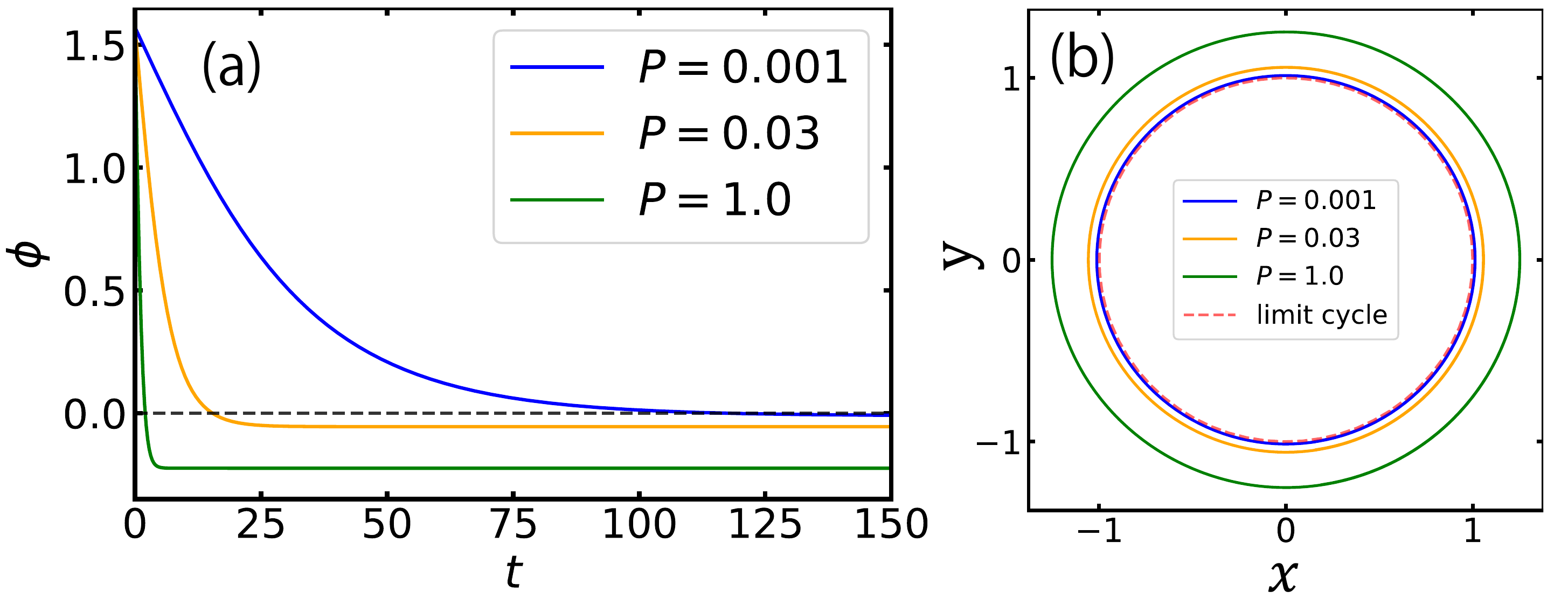}
    \caption{Results of phase-only optimization for the linear stability. (a)~Evolution of the phase differences. (b)~Trajectories on the phase plane in the steady state for input powers %$P=1.0\times10^{-3}$ (blue), $1.0\times10^{-2}$ (yellow), and $1.0\times10^{-1}$(green). 
    $P=0.001$ (blue), $0.03$ (yellow), and $1.0$ (green). 
    Black dashed line in (a) indicates the target phase difference $\phi^*=0$, and the red dashed curve in (b) shows the unperturbed limit cycle with $P=0$.}
    \label{fig_1} 
    \end{center}
\end{figure}

We use the Stuart-Landau oscillator, a normal form of the supercritical Hopf bifurcation~\cite{kuramoto1984chemical, nakao2016phase}, as an example, given by
\begin{align}
  \begin{pmatrix}
    \dot{x} \\ \dot{y}
  \end{pmatrix}
  = 
  \begin{pmatrix}
    x - a  y - (x-b  y)(x^2+y^2)\\ a x + y - ( b x + y)(x^2+y^2)
  \end{pmatrix}.
\end{align}
We assume the parameter values to be  $a = 11$ and $b = 1$, which gives the natural frequency $\omega = a-b = 10$
and non-zero Floquet exponent $ \lambda_1 = -2$. The input frequency is assumed to be equal to the natural frequency, $\Omega = 10$. The target phase-locking point is set as $\phi^* = 0$ and the initial condition of $(x, y)$ is chosen such that the phase difference at $t=0$ is $\phi(0) = \pi/2$.

Figures~\ref{fig_1}(a) and (b) show the evolution of the phase difference $\phi(t)$ and the trajectory on the $(x, y)$ phase plane, respectively, for three values of $P$. From Fig.~\ref{fig_1}(a), we observe that the convergence becomes faster as $P$ is increased, but the realized phase-locking point deviates more significantly from the target point $\phi^*=0$.
This discrepancy is caused by the breakdown of the phase-only approximation. As shown in Fig.~\ref{fig_1}(b), as the intensity of the input $P$ increases, the system state deviates from the unperturbed limit cycle more significantly, yielding larger errors in the phase reduction and invalidating the results of the optimization
that relies only on the phase equation.

%%%%%%%%%%%%%%%%%
%%% Section 4 %%%
%%%%%%%%%%%%%%%%%

\section{Optimal entrainment with amplitude suppression}

From the results in the previous section, it is expected that
faster convergence to the target phase-locking point can be realized if the amplitude deviations of the system state from the limit cycle can be suppressed.
To accomplish this, we propose two methods: an amplitude-feedback method and an amplitude-penalty method.

%%%%%%%%%%%%%%%%%%%%%%%%%%%%%%%%%%%%%%%%%%%%%%%%
\subsection{Amplitude-feedback method}
%%%%%%%%%%%%%%%%%%%%%%%%%%%%%%%%%%%%%%%%%%%%%%%%

The first method is to add an amplitude-feedback term to the optimal input ${\bm q}^{ls}$ derived in the previous section. We consider a modified periodic input given by
\begin{align}
  \bm{{q}}^{fb}(t) &= \bm{q}^{ls}(\Omega t) - \alpha {\bm y}(t)
  \label{eq:qfb}
\end{align}
where ${\bm y}(t) = {\bm X} - {\bm \chi}(\theta(t))$ represents the deviation of the system state from the limit cycle $\chi$ and $\alpha > 0$ is a feedback gain.
When ${\bm y}(t)$ is small, we can approximately expand it as
\begin{align}
  {\bm y}(t) \approx \sum_{i=1}^{M} r_i(t){\bm{u}}_i(\theta(t)/\omega)
\end{align}
by using the first $M$ amplitude variables $r_i$ for $i=1, ..., M$ ($M \leq N-1$). Therefore, the approximate phase and amplitude equations become
\begin{align}
  \dot{\theta}(t) 
  &= \omega +  \la \bm{Z}(\theta(t)),\ \bm{q}^{fb}(t) \ra 
  \cr
  &= \omega + \la \bm{Z}(\theta(t)),\ \bm{q}^{ls}(\Omega t) \ra
\end{align}
and
\begin{align}
  \dot{r}_i(t) 
  &= \lambda_i r_i(t) +  \la \bm{I}_i(\theta(t)),\ \bm{q}^{fb}(t) \ra
  \cr
  &= (\lambda_i - \alpha ) r_i(t) + \la \bm{I}_i(\theta(t)),\ \bm{q}^{ls}(\Omega t) \ra
\end{align}
from the bi-orthonormality relation of the left Floquet eigenvectors (PSF and ISFs) and the right Floquet eigenvectors, $\la {\bm Z}(\theta), {\bm u}_i(\theta/\omega) \ra = 0$ and $\la {\bm I}_i(\theta), {\bm u}_j(\theta/\omega) \ra = {\delta}_{ij}$ for $i,j=1, ..., M$.
Namely, at the lowest-order phase-amplitude reduction, the phase dynamics is unaffected by the feedback while the convergence rate of the amplitude $r_i$ changes from $\lambda_i$ to $\lambda_i - \alpha$ when the deviation ${\bm y}(t)$ is small.
We thus expect faster decay of the amplitudes and suppression of the deviation of the system state from the unperturbed limit cycle $\chi$.

In the numerical simulations described in the next section, a moderate value of the feedback gain $\alpha$ is chosen empirically so that the feedback input is sufficiently large to suppress the amplitude deviations but not too large to lose efficiency.
%

%%%%%%%%%%%%%%%%%%%%%%%%%%%%%%%%%%%%%%%%%%%%%%%%
\subsection{Amplitude-penalty method}
%%%%%%%%%%%%%%%%%%%%%%%%%%%%%%%%%%%%%%%%%%%%%%%%

The second method, which is feedforward, is to add a penalty term to the optimization problem, Eq.~(\ref{eq:opt_lins}), which increases if the amplitude deviations of the system state from the limit cycle are excited. Specifically, we consider the following optimization problem:
\begin{align}
  \label{eq:opt_pnt}
  &\max_{\bm{q}}\ -\Gamma'(\phi^*) - \sum_{i=1}^{M} k_i \left[  \la \bm{I}_i(\phi^*+\Omega t),\  \bm{q}(\Omega t) \ra ^2 \right]_t,
  \cr
  & \mbox{s.t.}
  \quad \Delta + \Gamma(\phi^*) = 0, ~ [ \| \bm{q} \|^2  ]_t = P,
\end{align}
where the second term in the objective function represents the penalty and
$k_i \geq 0 $ gives the weight of the penalty for the excitation of the $i$th amplitude.
We use the same weights $k_{i} = k_{i+1}$ if the $i$th and $i+1$th amplitudes are mutually complex conjugate, namely, when $\lambda_i = \lambda_{i+1}^*$.

The inner product $\la \bm{I}_i(\phi^*+\Omega t),\ \bm{q}(\Omega t) \ra$ is nothing but the second perturbation term in the amplitude equation~(\ref{eq:phi-r-lowest}) for the case that the oscillator phase is $\theta = \phi^* + \Omega t$ and the input phase is $\Omega t$; namely, the effect that the periodic input exerts on the $i$th amplitude when the oscillator is phase-locked to the periodic input at the target phase difference $\phi^*$.
By adding this term to the objective function, the functional component in the periodic input that causes amplitude deviations of the oscillator state from the limit cycle is expected to be diminished.

To solve the optimization problem Eq.~(\ref{eq:opt_pnt}), we introduce a Lagrange function
\begin{align}
  C(\bm{q}, \mu, \nu) = -\Gamma'(\phi^*) - \sum_{i=1}^{M} k_i [  \la \bm{I}_i(\phi^*+\Omega t),  \bm{q}(\Omega t) \ra ^2 ]_t
  + \mu ( \Delta + \Gamma(\phi^*)) + \nu (P - [ \| \bm{q} \|^2  ]_t ),
\end{align}
where $\mu$ and $\nu$ are the Lagrange multipliers and the amplitudes up to $M$ ($\leq N-1$) are taken into account. The conditions for the extremum yield
\begin{align}
  \label{eq:pena1}
  \frac{{\delta}{C}}{{\delta}{\bm{q}}} =& \frac{1}{T_e}\Big\{-\bm{Z'}(\phi^*+\Omega t) -2 \sum_{i=1}^{M} k_i \bm{I}_i(\phi^*+\Omega t) {\bm{I}_i}^\dag (\phi^*+\Omega t) \bm{q}(\Omega t) \cr
  &+ \mu \bm{Z}(\phi^*+\Omega t) - 2\nu \bm{q}(\Omega t) \Big\} = 0,
  \\
  \label{eq:pena2}
  \frac{\partial{C}}{\partial{\mu}} =& \Delta + [\la \bm{Z}(\phi^*+\Omega t) , \bm{q}(\Omega t) \ra]_t = 0,
  \\
  \label{eq:pena3}
  \frac{\partial{C}}{\partial{\nu}} =& P - [ \| \bm{q}(\Omega t)\|^2 ]_t = 0.
\end{align}
From Eq.~(\ref{eq:pena1}), we obtain
\begin{align}
  \label{eq:pena4}
  \bm{q}(\Omega t) = \frac{1}{2} \left\{\nu \bm{E} + \sum_{i=1}^{M} k_i \bm{I}_i(\phi^*+\Omega t) {\bm{I}_i}^\dag (\phi^*+\Omega t) \right\}^{-1} \left\{ -\bm{Z'}(\phi^*+\Omega t)+\mu\bm{Z}(\phi^*+\Omega t) \right\},
\end{align}
where $\bm{E} \in \mathbb{R}^{N\times N}$ is an identity matrix.
By plugging Eq.~(\ref{eq:pena4}) into Eq.~(\ref{eq:pena2}), we obtain
\begin{align}
  \Delta + \left[ \la \bm{Z} ,  \frac{1}{2} \left\{ \nu {\bm E} + \sum_{j=1}^{M} k_i \bm{I}_i{\bm{I}_i}^\dag \right\}^{-1} \left\{-\bm{Z'} +\mu\bm{Z}\right\} \ra
    \right]_t= 0,
\end{align}
from which the relation between $\mu$ and $\nu$ is derived as
\begin{align}
  \mu = \frac{ \left[ \la \bm{Z}, \{\nu \bm{E} + \sum_{i=1}^{M} k_i \bm{I}_i{\bm{I}_i}^\dag \}^{-1}\bm{Z'}\ra \right]_t - 2 \Delta}{
    \left[ \la \bm{Z} ,  \{\nu  \bm{E} + \sum_{i=1}^{M} k_i \bm{I}_i{\bm{I}_i}^\dag \}^{-1}\bm{Z}\ra \right]_t}. 
  \label{eq:pena_mu}
\end{align}
Though it is difficult to determine $\mu$ and $\nu$ analytically, we can numerically find an appropriate Lagrange multiplier $\nu$, which determines ${\bm q}$ and $\mu$  from Eqs.~(\ref{eq:pena4}) and~(\ref{eq:pena_mu}), such that the power constraint in Eq.~(\ref{eq:pena3}) is satisfied. We denote the resulting optimal input as $\bm{q}^{pl}$.

If we consider the limit $k_i \to \infty$ ($i=1, ..., M$) in Eq.~(\ref{eq:opt_pnt}), it is expected that the input waveform ${\bm q}$ can possess only the tangential component along the limit cycle because ${\bm q}$ should not excite the amplitude deviations at all.
In Appendix B, we present the result of optimization for such a tangential input and compare it with the results of the amplitude-penalty method. 

In the numerical simulations described in the next section, moderate values of the weights $k_i$ are chosen empirically so that the effect of the input exerting on the $i$th amplitude is sufficiently small and at the same time the penalty does not degrade the improvement in the linear stability.
%

%%%%%%%%%%%%%%%%%
%%% Section 5 %%%
%%%%%%%%%%%%%%%%%
\section{Examples}

In this section, we illustrate the results of the two methods using a two-dimensional van der Pol model~\cite{van1927frequency,van1927vii} with a real non-zero Floquet exponent and a three-dimensional Willamowski-R\"ossler model~\cite{willamowski1980irregular} that possesses a pair of complex non-zero Floquet exponents.

\begin{figure}[!t]
  \includegraphics[width=0.8\hsize,clip]{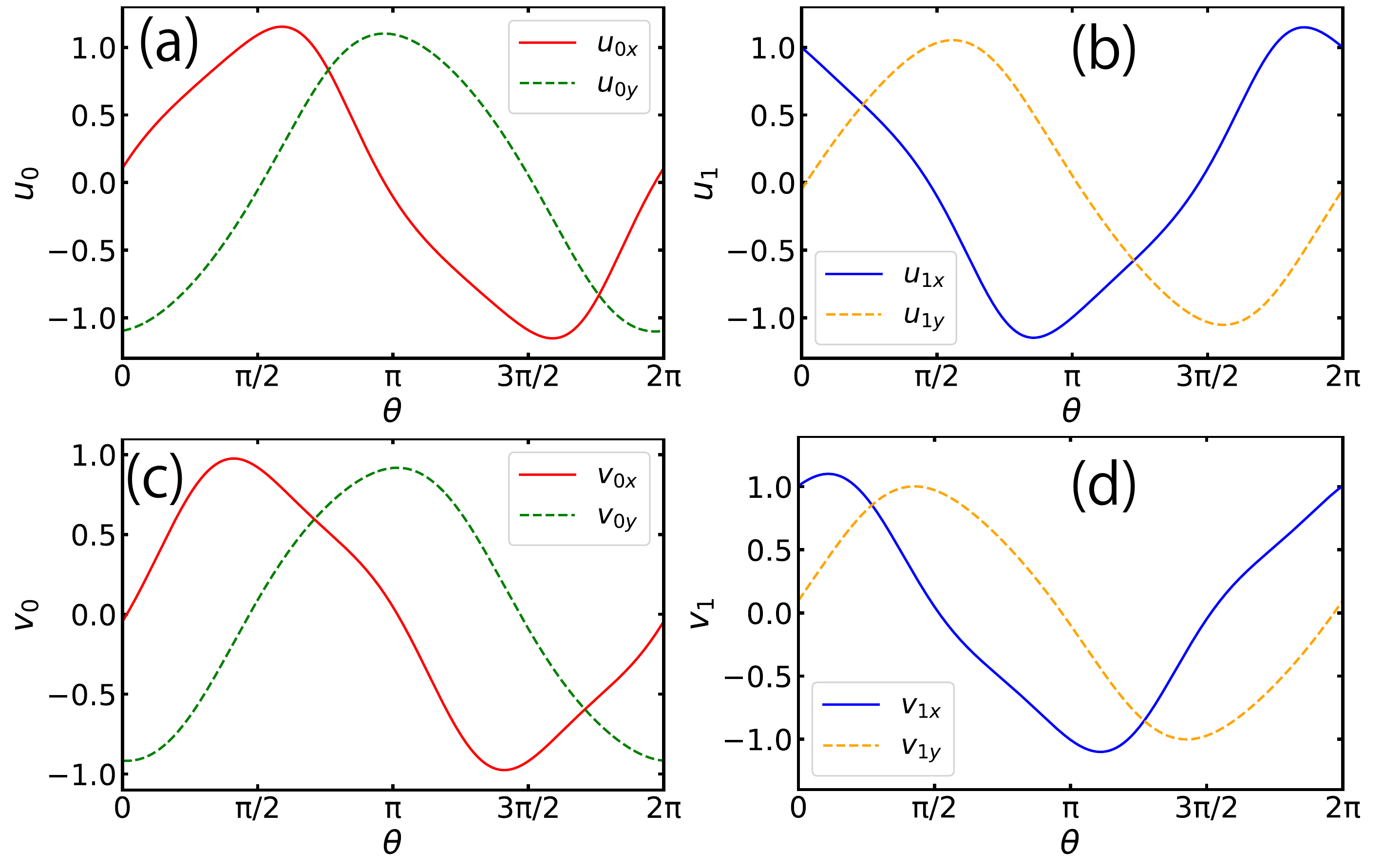}
  \caption{Floquet eigenvectors $\bm{u}_i(\theta/\omega), \bm{v}_i(\theta/\omega)$ of the van der Pol model for $i=0$ (a,c) and $i=1$ (b, d). Both $x$ and $y$ components of the eigenvector are shown in each figure.}
  \label{fig_2}
\end{figure}

%%%%%%%%%%%%%%%%%%%%%%%%%%%%%%%%%%%%%%%%%%%%%%%%
\subsection{van der Pol model}
%%%%%%%%%%%%%%%%%%%%%%%%%%%%%%%%%%%%%%%%%%%%%%%%

We consider the van der Pol model of an electric oscillator~\cite{van1927frequency,van1927vii}
given by
\begin{align}
  \begin{pmatrix}
    \dot{x} \\ \dot{y}
  \end{pmatrix}
  = 
  \begin{pmatrix}
    d(c x - x^3/3 - y)\\ dx
  \end{pmatrix},
\end{align}
where $x$ and $y$ represent the current and voltage, respectively, and $c$ and $d$ are parameters, which we fix as $c = 0.3$ and  $d = 10$.
The natural frequency of the limit cycle is $\omega \approx 9.94$ and the non-zero Floquet exponent is $\lambda_1 \approx -3.02$. 
The Floquet eigenvectors are plotted in Fig.~\ref{fig_2} (See Appendix B for the numerical methods).
We fix the input power as 
$P=1.0$
and the input frequency as 
$\Omega = \omega + 0.5 \approx  10.44$.
The value $P=1.0$ is already non-weak for this system and can excite amplitude deviations if neither feedback nor penalty method is introduced.
The target phase-locking point is $\phi^* = 0$ and the initial system state $(x, y)$ at $t=0$ is chosen such that the initial phase difference is $\phi(0) = \pi/2$.
The feedback gain in the amplitude-feedback method is  
$\alpha = 50$ and the weight in the amplitude-penalty method is $k_1=20$ (note that $N=2$).

\begin{figure}[!t]
  \includegraphics[width=0.8\hsize,clip]{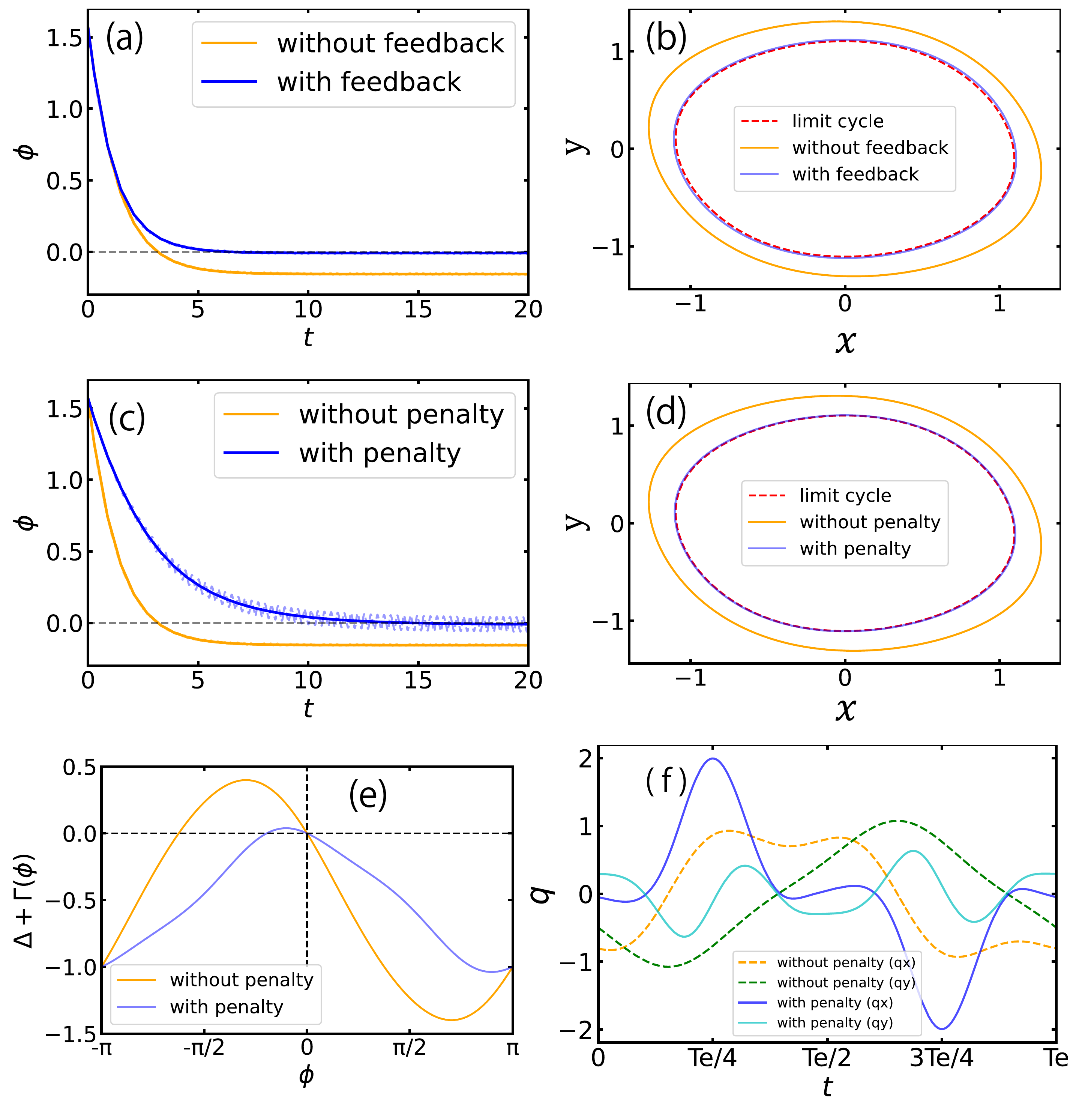}
  \caption{Results of amplitude-feedback and amplitude-penalty methods for the van der Pol model.
    (a) Evolution of the phase differences (raw and averaged) and (b) Trajectory on the phase plane for the amplitude-feedback method.
    (c) Evolution of the phase differences (raw and averaged) and (d) Trajectory on the phase plane for the amplitude-penalty method.
    (e) $\Delta + \Gamma(\phi)$ in Eq.~(\ref{phiT}) for the cases with and without the amplitude penalty.
    (f) Optimal inputs with and without amplitude penalty. 
	In (a) and (c), the black dashed lines represent the target phase-locking point $\phi^* = 0$.
    In (b) and (d), the red dashed curves show the unperturbed limit cycle when $P=0$.
    }
  \label{fig_3}
\end{figure}

The results of the optimal entrainment with the amplitude-feedback method are shown in Figs.~\ref{fig_3}(a) and (b). The optimal input waveforms in this case are given by the dashed curves ('without penalty') in Fig.~\ref{fig_3}(f).
The phase difference $\phi$ converges to the correct target value $\phi^*=0$ when the amplitude feedback is applied.
This is because the deviation of the system state from the unperturbed limit cycle is suppressed as shown in Fig.~\ref{fig_3}(b) and the validity of the phase equation is kept.
In contrast, without the feedback, 
large amplitude deviations occur as shown in Fig.~\ref{fig_3}(b) and $\phi$ converges to an incorrect value.

The results of the optimal entrainment with the amplitude-penalty method are shown in Figs.~\ref{fig_3}(c), (d), (e) and (f).
The phase difference converges to the correct target value also in this case when the penalty is introduced.
This is because the optimal input with the amplitude penalty does not excite large amplitude deviations and the system state stays near the unperturbed limit cycle as shown in Fig.~\ref{fig_3}(d).
In contrast, for the optimal input without the penalty, 
large amplitude deviations occur and $\phi$ converges to an incorrect value.

Figure~\ref{fig_3}(e) shows $\Delta + \Gamma(\phi)$ in Eq.~(\ref{phiT}), calculated using the optimal inputs with and without the penalty plotted in Fig.~\ref{fig_3}(f).
The linear stability of the phase-locking point $\phi^* = 0$ is given by the (negative) slope of the curve at the origin. 
The case without the penalty gives higher stability than the case with the penalty because some of the vector components that are included in the input waveform without the penalty are dropped in the input waveform with the penalty to avoid excitation of the amplitude deviations.
Thus, the optimal input without penalty should lead to faster convergence when the input power $P$ is sufficiently weak.
However, because $P$ is not weak in this simulation, the input without the penalty yields large deviations of the system state from the limit cycle
and does not realizes the correct target value.

In Figs.~\ref{fig_3}(a,c), we plotted two values of the phase difference for each curve, that is, the raw values directly measured by using the phase function $\Theta$ and the averaged values calculated by taking a moving average of the raw values 
for one period $T_e = 2 \pi/\Omega$ of the periodic input, i.e., $\bar{\phi}(t) = [ \phi(t+s) ]_s$, where the latter values correspond to the averaged phase variable used in Eq.~(\ref{phiT}) (though we used the same symbol $\phi$ for both raw and averaged phase variables for simplicity).
The raw phase $\phi(t)$ naturally exhibits small wobbling around the averaged phase $\bar{\phi}(t)$. 

In this two-dimensional van der Pol model, the case with the penalty yielded larger wobbling than the case with the amplitude feedback. This is because the input waveforms with the penalty fluctuates more strongly than those without the penalty as shown in Fig.~\ref{fig_3}(f) in order not to perturb the oscillator when the ISFs shown in Fig.~\ref{fig_2}(d) take large values.
The wobbling of all curves can be well suppressed by taking the moving average.

\begin{figure}[!t]
  \includegraphics[width=0.85\hsize,clip]{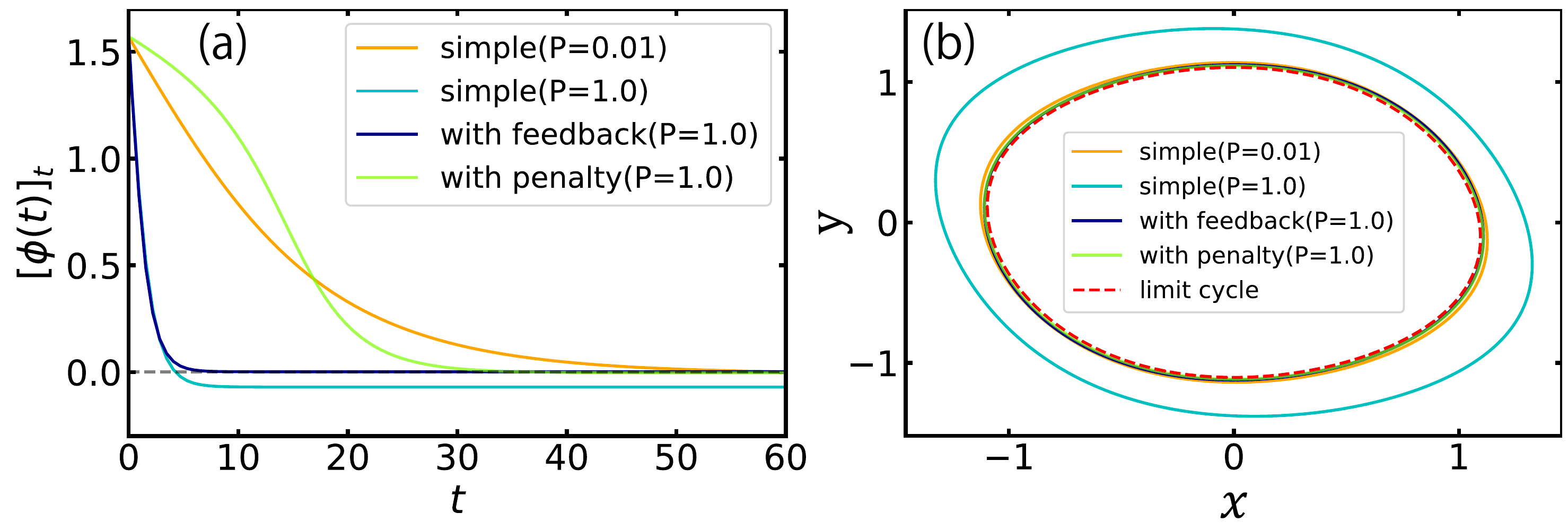}
  \caption{
    Optimal entrainment by the amplitude-feedback and amplitude-penalty methods for strong input ($P=1.0$). For comparison, the cases without feedback nor penalty are also shown for weak ($P=0.01$) and strong ($P=1.0$) inputs. (a) Evolution of the phase differences (averaged). (b) Trajectories on the phase plane in the steady state. 
  }
  \label{fig_4}
\end{figure}

Figure~\ref{fig_4} shows the optimal entrainment for the cases with the amplitude feedback 
($\alpha=50$) and with the amplitude penalty 
($k=10$), 
where the input power  $P=1.0$ is the same as that used in Fig.~\ref{fig_3}. 
For comparison, the results without the feedback nor the penalty are also shown as the simple cases for very weak ($P=0.01$) and strong ($P=1.0$) inputs.
Here we fix the input frequency as $\Omega = \omega \approx 9.94$.

As shown in Fig.~\ref{fig_4}(a), without the amplitude suppression, the phase difference converges to the correct target value $\phi^*=0$ only for the weak input ($P=0.01$) and fails to converge appropriately for the strong input ($P=1.0$). 
In contrast, both feedback and penalty methods achieve accurate convergence to the target phase-locking point for the strong input ($P=1.0$). 
Figure~\ref{fig_4}(b) shows that both methods suppress amplitude deviations of the system state from the limit cycle even under the strong input, while the system state largely deviates from the limit cycle for the strong input if neither feedback nor penalty are given.

It is remarkable that the proposed methods with the amplitude suppression can achieve faster convergence to the correct target phase-locking point than the conventional simple case without the amplitude suppression.
The feedback method can realize much faster convergence by using a strong feedback gain, while the improvement in the performance of the penalty method is rather moderate. However, it should be stressed that the penalty method gives feedforward control and does not require the measurement of the system state.

\begin{figure}[!t]
  \includegraphics[width=1.0\hsize,clip]{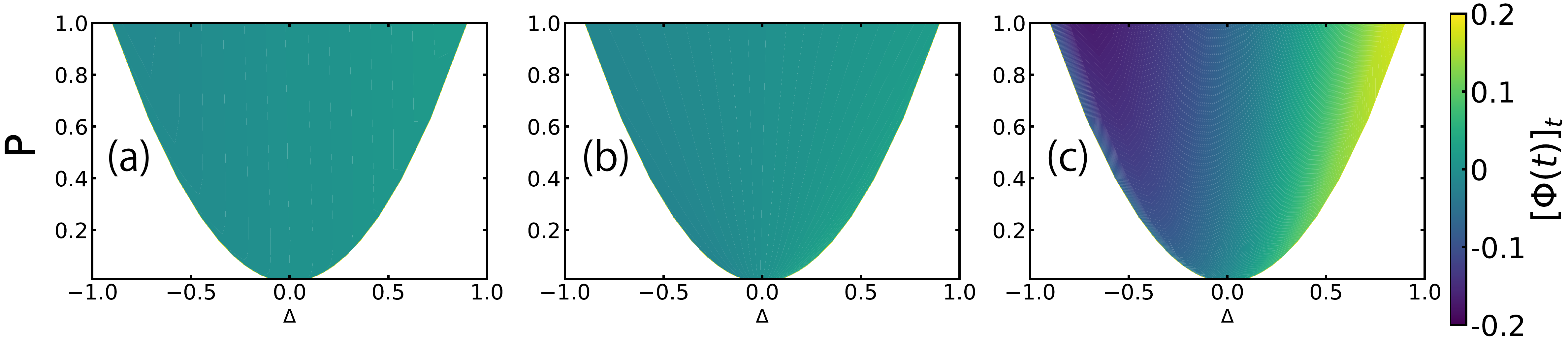}
  \caption{Arnold tongues and realized phase-locking points for (a) amplitude-feedback method, (b) amplitude-penalty method, and (c) no amplitude suppression. The color map represents the discrepancy between the target phase-locking point $\phi^*=0$ and the realized phase-locking point.
  }
  \label{fig_5}
\end{figure}

The amplitude-feedback and amplitude-penalty methods can also widen the parameter region in which the phase difference converges to the correct target phase-locking point.
Figure~\ref{fig_5} shows the Arnold tongues for the (a) amplitude-feedback method ($\alpha=50$), (b) amplitude-penalty method
($k=20$), 
and (c) without the amplitude suppression. In each figure, the horizontal axis represents the frequency mismatch $\Delta=\omega-\Omega$ and the vertical axis represents the input power $P$.
In each Arnold tongue, the discrepancy between the realized phase-locking point (averaged over one oscillation period after convergence) and the target phase-locking point $\phi^*=0$ is visualized by a color map.
Accurate convergence to $\phi^*=0$ is confirmed in wide parameter regions in Fig.~\ref{fig_5}(a) (amplitude-feedback) and Fig.~\ref{fig_5}(b) (amplitude-penalty). In contrast, in the case without the amplitude suppression in Fig.~\ref{fig_5}(c), the realized phase-locking point shows considerable deviations from $\phi^*=0$ for large $P$.

Thus, both proposed methods enable us to apply stronger periodic inputs by suppressing the amplitude deviations and achieve considerably faster entrainment and wider entrainment ranges. 

\subsection{Willamowski-R\"ossler model}

\begin{figure}[!t]
  \includegraphics[width=1\hsize,clip]{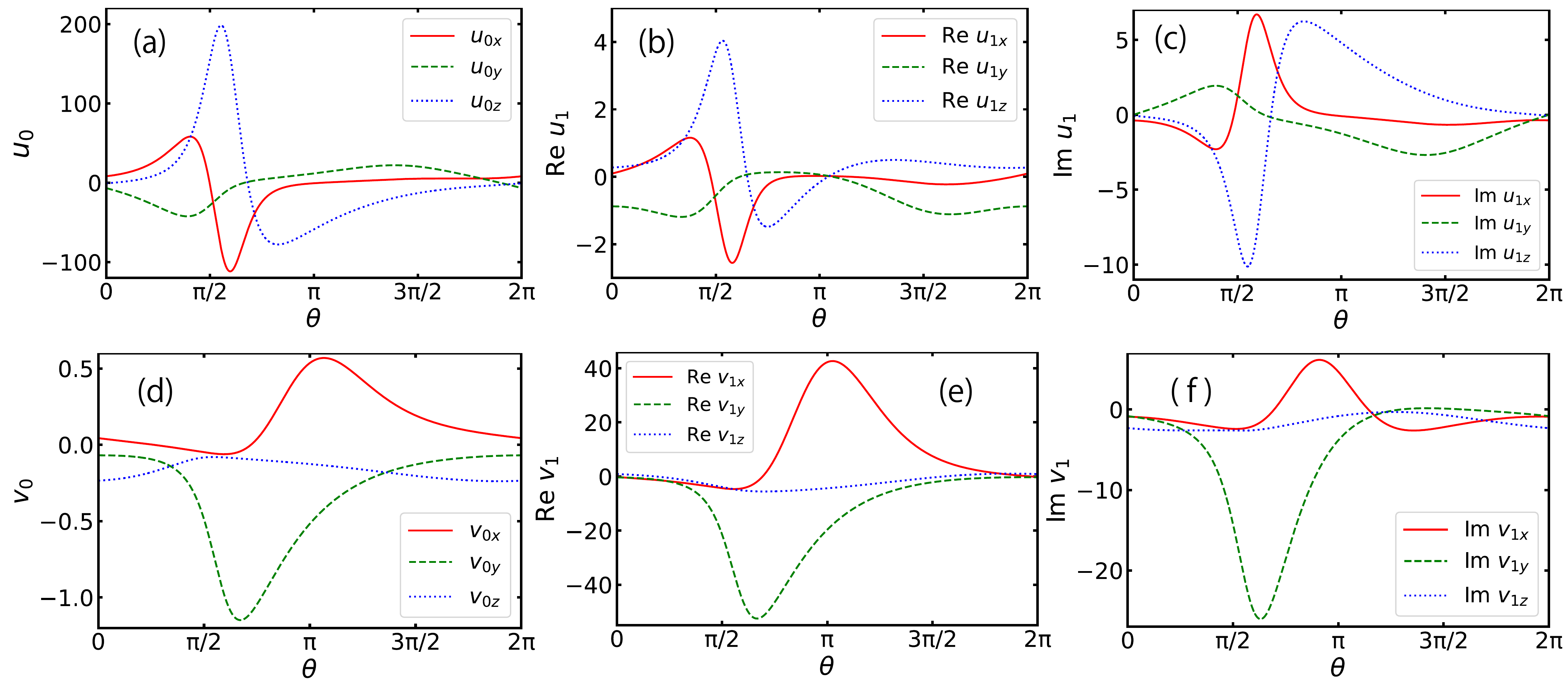} 
  \caption{Floquet eigenvectors $\bm{u}_i(\theta/\omega), \bm{v}_i(\theta/\omega)$ of the Willamowski-R\"ossler model for $i=0$ (a, d) and $i=1$ (b, e, c, f, real and imaginary parts). Note that the first Floquet eigenvectors are complex and the second Floquet eigenvectors are their complex conjugate, i.e., $\bm{u}_2 = {\bm{u}_1}^*$ and $\bm{v}_2 = {\bm{v}_1}^*$.}
  \label{fig_6}
\end{figure}

The Willamowski-R\"ossler model for chemical oscillations~\cite{willamowski1980irregular,boland2009limit} is described by (in the expression of Ref.~\cite{boland2009limit})
\begin{align}
  \frac{d}{dt} \left(
  \begin{array}{c}
    x_1\\
    x_2\\
    x_3
  \end{array}
  \right)
  =  
  \left(
  \begin{array}{c}
    x _1 (b_1 - d_1 x_1 - x_2 - x_3) \\
    x_2(b_2 - d_2 x_2 - x_1) \\
    x_3 ( x_1 - d_3)
  \end{array}
  \right),
\end{align}
where $x_1$, $x_2$, and $x_3$ represent the concentrations of intermediate chemical species and the parameters $b_1 , b_2, d_1, d_2$, and $d_3$ are reaction rates. This system has 6 fixed points~\cite{boland2009limit} and exhibits various dynamics including limit-cycle oscillations and chaos~\cite{geysermans1996particle,aguda1988dynamic}.
We set the parameters as $b_1 = 80$, $b_2 = 20$, $d_1 = 0.16$, $d_2 = 0.13$, and $d_3 = 16$, with which the system has a stable limit cycle. The frequency of this limit cycle is $\omega \approx 17.25$ and the non-zero Floquet exponents are given by a pair of complex values $\lambda_1, \lambda_2 \approx -3.280\pm4.326\sqrt{-1}$.
Figure~\ref{fig_6} shows the Floquet eigenvectors $i=0$ (real) and $i=1, 2$ (mutually complex conjugate) obtained numerically (see Appendix B for the numerical methods).

We set the input power and frequency as $P=10$ and $\Omega = \omega + 0.3 \approx  17.55$, respectively.
As shown below, this value of $P$ is considerably strong and the system state largely deviates from the unperturbed limit cycle if neither feedback nor penalty is introduced.
The target phase-locking point is $\phi^* = 0$ and the initial state is prepared such that the initial phase difference is $\phi(0) = \pi/2$.
We use a feedback gain $\alpha = 1.0\times10^3$ for the amplitude-feedback method and weights $k_1=k_2=0.1$ for the amplitude-penalty method ($N=3$).

\begin{figure}[!t]
  \includegraphics[width=0.8\hsize,clip]{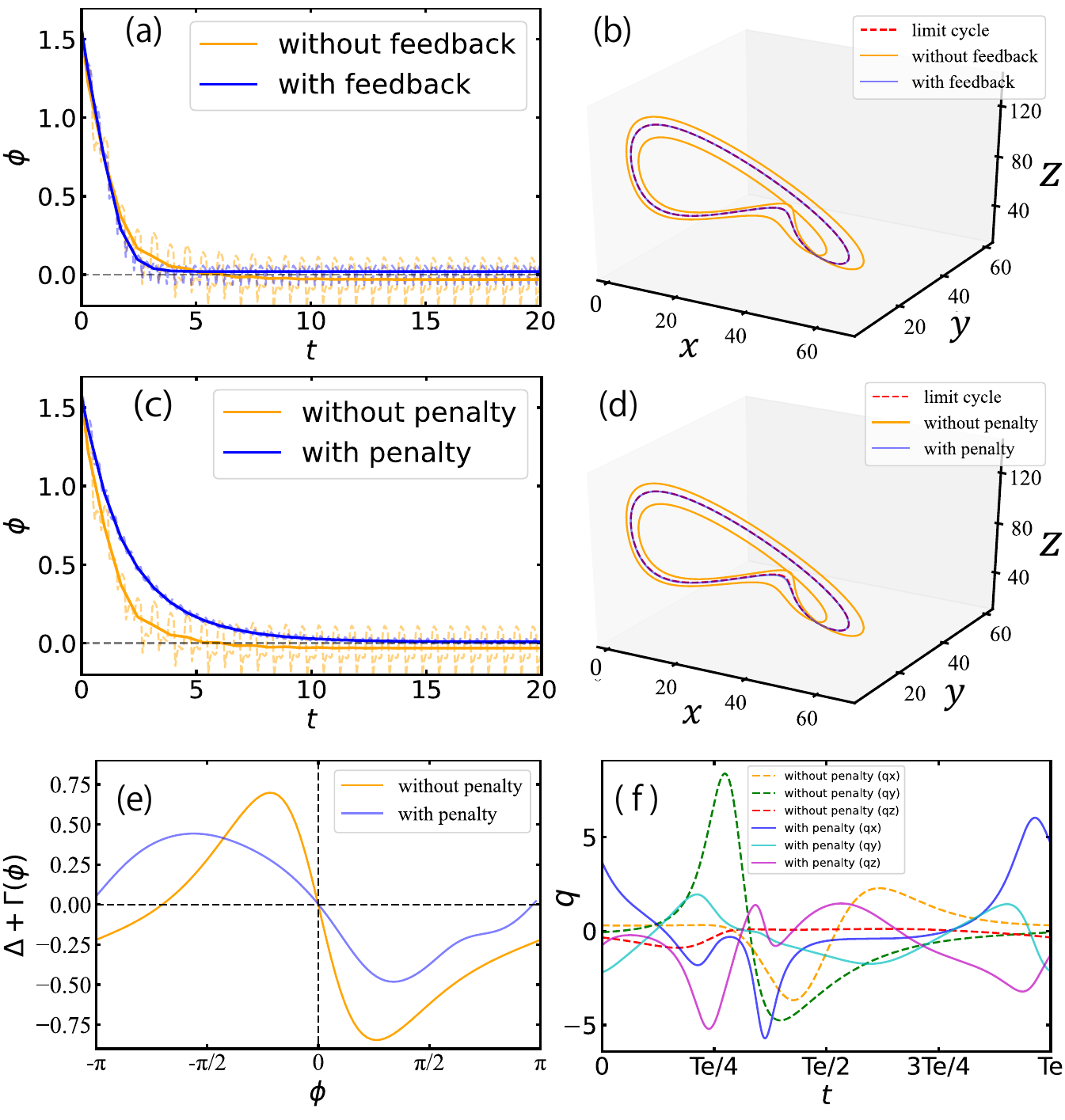}
  \caption{Results of amplitude-feedback and amplitude-penalty methods for the Willamowski-R\"ossler model.
    (a) Evolution of the phase differences (raw and averaged) and (b) Trajectories on the phase plane for the amplitude-feedback method.
    (c) Evolution of the phase differences (raw and averaged) and (d) Trajectories on the phase plane for the amplitude-penalty method. 
    	(e) $\Delta + \Gamma(\phi)$ in Eq.~(\ref{phiT}) for the cases with and without the amplitude penalty.
    	(f) Optimal inputs with and without the amplitude penalty. 
    In (a) and (c), the black dashed lines represent the target phase-locking point $\phi^*=0$. In (b) and (d), the red dotted lines represent the unperturbed limit cycle with $P=0$.
  }
  \label{fig_7}
\end{figure}

Figures~\ref{fig_7}(a) and (b) show the optimal entrainment with the amplitude-feedback method. The deviations of the system state from the unperturbed limit cycle is suppressed as shown in Fig.~\ref{fig_7}(b) and the phase difference $\phi$ converges to the correct target phase-locking point $\phi^*=0$
as shown in Fig.~\ref{fig_7}(a).
In contrast, without the feedback, the system state largely deviates from the limit cycle and the phase difference does not accurately converge to $\phi^*=0$.
Indeed, the oscillator is so strongly perturbed by the periodic input that the system undergoes a period-doubling bifurcation and exhibits a longer closed orbit different from the unperturbed limit cycle as shown in Fig.~\ref{fig_7}(b).

Figure~\ref{fig_7}(c) and (d) show the optimal entrainment with the amplitude-penalty method.
We can confirm in Fig.~\ref{fig_7}(c) that the amplitude deviations are suppressed and the phase difference converges to the correct target phase-locking point in the case with the penalty.
In contrast, as shown in Fig.~\ref{fig_7}(d), the case without the penalty induces large amplitude deviations, which causes strong fluctuations of the phase difference and failure in realizing the correct target value.
As in the previous case of the van der Pol model, the linear stability of $\phi^* = 0$ is higher in the case without penalty as shown in Fig.~\ref{fig_7}(e), where $\Delta + \Gamma(\phi)$ in Eq.~(\ref{phiT}) calculated by using the optimal inputs shown in Fig.~\ref{fig_7}(f) is plotted.
However, because $P$ is not small, the input without the penalty fails to converge correctly.

In this model, without the amplitude suppression, the raw phase differences exhibit strong wobbling as shown in Figs.~\ref{fig_7}(a) and (c). This is caused by the considerable deviations of the system state from the unperturbed limit cycle due to strong inputs as shown in Figs.~\ref{fig_7}(b) and (d).
Both the amplitude-feedback and amplitude-penalty methods suppress the wobbling in the phase differences in this model, in contrast to the case of the van der Pol model.
This is because suppression of the amplitude deviations does not necessarily cause fluctuations in the phase direction in this three-dimensional model.
In Figs.~\ref{fig_7}(a) and (c), we averaged the raw phases for two periods $2T_e$ of the input because the system without feedback nor penalty exhibits a period-doubled orbit due to the strong effect of the periodic input as shown in Figs.~\ref{fig_7}(b) and (d).
%

%%%%%%%%%%%%%%%%%
%%% Section 6 %%%
%%%%%%%%%%%%%%%%%

\section{Concluding remarks}

In this paper, based on the phase-amplitude reduction, 
we have proposed two methods to derive strong input waveforms that suppress amplitude deviations and realize fast entrainment of limit-cycle oscillators,
namely, the amplitude-feedback and amplitude-penalty methods.
We used the van der Pol and Willamowski-R\"ossler models as examples and demonstrated that both methods enable us to apply stronger inputs while avoiding the breakdown of the phase equation, thereby realizing faster and accurate convergence to the target phase-locking point and wider locking range than the conventional method without the amplitude suppression.

The amplitude-feedback method can realize much faster entrainment by using a large feedback gain, but it assumes that we can always obtain the system state and the phase value, which could be difficult in practical implementation. Development of efficient methods to estimate the system's phase from a small number of measurements would thus be desirable.
Regarding the amplitude-penalty method, it can realize moderately faster entrainment in the feedforward setting, i.e., without measuring the system state. However, because it considers only the local amplitude deviations at the target phase-locking point, it only guarantees that the system state is close to the unperturbed limit cycle in the steady entrained state and not in the transient. It would thus be desirable to devise a method that can suppress also the amplitude deviations in the whole transient process for a more general, global optimization.

In the penalty method, we used a penalty term of the form
$\left[  \la \bm{I}_i(\phi^*+\Omega t),\  \bm{q}(\Omega t) \ra ^2 \right]_t$
in the objective function in Eq.~(\ref{eq:opt_pnt}) in order to suppress the effect of the input ${\bm q}(\Omega t)$ exerted on the amplitude $r_i(t)$.
This form of the penalty allowed us to derive the analytical solution Eq.~(\ref{eq:pena4}) to the optimization problem.
Instead of the above form, it may also be possible to include a more direct penalty term, namely, the squared mean of the amplitudes $ [\abs{r_i(t)}^2]_t$ in order to suppress the amplitude deviations caused by ${\bm q}(\Omega t)$.
However, the optimization problem with this penalty term is difficult to solve analytically and it should be solved numerically as a nonlinear programming problem (e.g. Ref.~\cite{kato2021optimization}).
In this study, we chose the former one for the analytical tractability and confirmed that it reasonably suppressed the amplitude deviations and achieved faster entrainment in numerical simulations.

It should also be mentioned that there are other approaches that allow us to apply strong inputs for entrainment or synchronization by adaptively modifying the response curves to take into account the changes caused by strong perturbations~\cite{kurebayashi2013phase, cui2009functional, castejon2013phase, rosenblum2019numerical, wilson2021optimal}.
In comparison to these adaptive methods, our present approach requires only the fundamental information near the limit cycle, namely, the PSF and ISFs that can be  evaluated simply by solving the adjoint equations, and hence it would be more easily implemented when the underlying model is known.
It would also be possible to generalize the present method to consider the effect of uncertainties in these response functions as in Refs.~\cite{wilson2015optimal, wilson2014energy}.

An important future topic is to develop a method to realize the present method of fast entrainment in a fully data-driven way. 
In the present study, we assumed that the system model is known, but it often happens in practical applications that the underlying dynamics is unknown.
In such cases, data-driven methods as discussed, e.g., in Refs.~\cite{monga2020supervised, wilson2020data, wilson2021data} should be necessary.
Because only the PSF and ISFs are required in the present methods, it may be possible to devise a simple algorithm to infer these quantities from observed data.

The methods proposed in the present study can also be used in different types of problems and allow us to apply stronger inputs to the oscillator for synchronization and entrainment.
For example, we will be able to use the present method for minimizing the control power of oscillators~\cite{moehlis2006optimal, dasanayake2011optimal, zlotnik2012optimal, li2013control} or maximizing the locking range of periodically driven oscillators~\cite{harada2010optimal, tanaka2014optimal, tanaka2015optimal}.
We will also be able to use the present methods for optimizing mutual synchronization of coupled oscillators.
Moreover, though we considered only the optimization of the local linear stability of the phase-locked state, we may be able to apply the present methods to other optimization problems, such as those for global entrainment property~\cite{kato2021optimization} and phase-distribution control~\cite{monga2018synchronizing, kuritz2019ensemble, monga2019phase2, kato2021optimization}.
Finally, using the phase-amplitude reduction frameworks for time-delayed~\cite{kotani2020nonlinear} and spatially-extended~\cite{nakao2021phaseamplitude} systems, we would also be able to realize fast entrainment in such non-conventional infinite-dimensional systems via amplitude suppression.
We thus expect broad applicability of the proposed methods in the control of limit-cycling dynamics using strong perturbations.
\\

\begin{acknowledgements}
We thank anonymous reviewers for the insightful comments, which significantly helped us improve the quality of this article.
This research was funded by JSPS KAKENHI JP17H03279, JP18H03287, JPJSBP120202201, JP20J13778, and JST CREST JP-MJCR1913.
\end{acknowledgements}

\subsection*{Data availability}

%The data that supports the findings of this study are available within the article
The source codes used for generating the data and figures in this paper are publicly available on GitHub
at \texttt{https://github.com/Shohei-Takata/Fast-optimal-entrainment}~\cite{github}.

\appendix

%%%%%%%%%%%%%%%%%
%%% Appendix 0 %%
%%%%%%%%%%%%%%%%%

\section{Derivation of the phase-amplitude equations for non-weak inputs}

In this section, we derive the first-order approximate phase-amplitude equations~(\ref{eq:pr_phi_r}).
We denote the right and left Floquet eigenvectors in Eq.~(\ref{floqueteigenvectors}) as
\begin{align}
{\bm U}_i(\theta) = {\bm u}_i(\theta / \omega),
\quad
{\bm V}_i(\theta) = {\bm v}_i(\theta / \omega),
\end{align}
for simplicity. From Eq.~(\ref{eq:adj_eq}), these eigenvectors satisfy 
\begin{align}
\frac{d {\bm U}_i(\theta)}{d\theta} = \frac{ 1 }{\omega} [{\bm J}(\theta) - \lambda_i] {\bm U}_i(\theta),
\quad
\frac{d {\bm V}_i(\theta)}{d\theta} = - \frac{ 1 }{\omega} [ {\bm J}(\theta)^{\dag}- \lambda_i^*  ]{\bm V}_i(\theta),
\end{align}
where ${\bm J}(\theta) = {\bm J}({\bm \chi}(\theta)) = {\bm J}({\bm X}_0(t=\theta/\omega))$ is the Jacobian matrix of ${\bm F}({\bm X})$ at ${\bm X} = {\bm \chi}(\theta)$ and $\lambda_i$ is the $i$th Floquet exponent.
In particular, ${\bm U}_0(\theta)$ is tangent to the limit-cycle orbit ${\bm \chi}(\theta)$ and we choose it as
\begin{align}
{\bm U}_0(\theta) = \frac{d{\bm \chi}(\theta)}{d\theta} = \frac{dt}{d\theta} \frac{d {\bm \chi}(\theta = \omega t)}{dt} = \frac{1}{\omega} {\bm F}({\bm \chi}(\theta)).
\end{align}
These eigenvectors satisfy the bi-orthogonality relation in the main text, i.e., 
\begin{align}
\la {\bm V}_j(\theta),\ {\bm U}_k(\theta) \ra = \delta_{j,k}.
\end{align}
Note that ${\bm Z}(\theta) = {\bm V}_0(\theta)$ and ${\bm I}_i(\theta) = {\bm V}_i(\theta)$ ($i=1, ... N-1$).

We consider a perturbed limit-cycle oscillator given by (Eq.~(\ref{eq:xperturb}) in the main text)
\begin{align}
	\dot{\bm X}(t) = \bm{F}({\bm X}(t)) + \bm{p}(t),
	\label{A0}
\end{align}
where we do not assume ${\bm p}(t)$ to be weak. Instead, we assume that  the system state ${\bm X}$ is kept close to the limit cycle within the distance of $\mathcal{O}(\delta)$ by appropriately choosing the functional form of ${\bm p}$.

As explained in the main text, we introduce the phase function $\Theta$, define the phase variable of the oscillator state ${\bm X}$ as $\theta = \Theta({\bm X})$, and represent the oscillator state ${\bm X}$ as
\begin{align}
{\bm X}(t) = {\bm \chi}(\theta(t)) + {\bm y}(t),
\label{A1}
\end{align}
where ${\bm y}(t)$ is a deviation of ${\bm X}(t)$ from the state ${\bm \chi}(\theta(t))$ on the limit cycle with the same phase $\theta(t)$ as ${\bm X}(t)$.
We expand ${\bm y}(t)$ using right Floquet vectors as
\begin{align}
{\bm y}(t) = \sum_{j=1}^{N-1} c_j (t) {\bm U}_j(\theta(t)),
\label{A2}
\end{align}
where $ c_j (t) = \langle {\bm V}_j(\theta),\ {\bm y}(t) \rangle$ is the $j$th expansion coefficient of $\mathcal{O}(\delta)$.
These expansion coefficients are approximately equal to the amplitudes, because $R_j({\bm X}(t)) = R_j({\bm \chi}(\theta(t)) + {\bm y}(t)) = R_j({\bm \chi}(\theta(t))) + \langle \nabla R_j({\bm \chi}(\theta(t))),\ {\bm y}(t) \rangle + \mathcal{O}(\delta^2) = \langle {\bm I}_j(\theta(t)),\ {\bm y}(t) \rangle +  \mathcal{O}(\delta^2)$, where we used $R_j({\bm \chi}(\theta(t))) = 0$ and ${\bm V}_j(\theta) = {\bm I}_j(\theta) = \nabla R_j({\bm \chi}(\theta))$.
Thus, we have $r_j(t) = R_j({\bm X}(t)) = c_j(t) + O(\delta^2)$ and $r_j(t) = \mathcal{O}(\delta)$ $(j=1, ..., N-1)$.

Plugging Eqs.~(\ref{A1}) and ~(\ref{A2}), and the above expression into Eq.~(\ref{A0}), we have
\begin{align}
\frac{d {\bm X}}{dt} 
&= \frac{d {\bm \chi}(\theta)}{dt} + \frac{d {\bm y}}{dt} 
\cr
&= \frac{d\theta}{dt} \frac{d{\bm \chi}(\theta)}{d\theta} + \sum_j \frac{d r_j}{dt} {\bm U}_j(\theta) + \sum_j r_j \frac{d\theta}{dt} \frac{d{\bm U}_j(\theta)}{d\theta} + O(\delta^2)
\cr
&= \frac{d\theta}{dt}  {\bm U}_0(\theta) + \sum_j \frac{d r_j}{dt} {\bm U}_j(\theta) + \sum_j r_j \frac{d\theta}{dt} \frac{ 1 }{\omega} [ {\bm J}(\theta) - \lambda_j ] {\bm U}_j(\theta) + O(\delta^2)
\end{align}
for the left-hand side and
\begin{align}
{\bm F}({\bm X}) + {\bm p} 
&= {\bm F}({\bm \chi}(\theta) + {\bm y}) + {\bm p} 
\cr
&= {\bm F}({\bm \chi}(\theta)) + {\bm J}(\theta) {\bm y} +
\mathcal{O}(\delta^2) + {\bm p}
\cr
&= {\omega} {\bm U}_0(\theta) + \sum_j r_j {\bm J}(\theta) {\bm U}_j(\theta) + {\bm p} + 
\mathcal{O}(\delta^2)
\end{align}
for the right-hand side. Therefore, 
\begin{align}
\frac{d\theta}{dt} \left( {\bm U}_0(\theta)  + \sum_j r_j \frac{1}{\omega} \left[ {\bm J}(\theta) - \lambda_j \right] {\bm U}_j(\theta) \right) + \sum_j \frac{d r_j}{dt} {\bm U}_j(\theta) 
\cr
= 
{\omega} {\bm U}_0(\theta) + \sum_j r_j {\bm J}(\theta) {\bm U}_j(\theta) + {\bm p} 
+
\mathcal{O}(\delta^2).
\label{A9}
\end{align}
We  assume that the last $\mathcal{O}(\delta^2)$ term is sufficiently small and drop it in the following analysis.

First, taking an inner product of Eq.~(\ref{A9}) with ${\bm V}_0(\theta)$, we obtain
\begin{align}
\frac{d\theta}{dt} \left( 1 + \frac{1}{\omega} \sum_j r_j \langle {\bm V}_0(\theta),\ {\bm J}(\theta) {\bm U}_j(\theta) \rangle \right) 
=
\omega + \sum_j r_j \langle {\bm V}_0(\theta),\ {\bm J}(\theta) {\bm U}_j(\theta) \rangle
+ 
\langle {\bm V}_0(\theta),\ {\bm p} \rangle 
\end{align}
and therefore the equation for the phase $\theta$ is given by
\begin{align}
\frac{d\theta}{dt} 
&= \omega + \frac{ \langle {\bm V}_0(\theta),\ {\bm p} \rangle  }{ 1 + \frac{1}{\omega} \sum_j r_j \langle {\bm V}_0(\theta),\ {\bm J}(\theta) {\bm U}_j(\theta) \rangle }
=  \omega + \langle {\bm V}_0(\theta),\ {\bm p} \rangle
+ \mathcal{O}\left( \frac{ \delta }{\omega} \right),
\label{A12}
\end{align}
where we expanded the denominator and used that $r_j = \mathcal{O}(\delta)$ in the second expression. Thus, the phase equation is expressed as
\begin{align}
\frac{d\theta}{dt} 
=  \omega + \langle {\bm V}_0(\theta),\ {\bm p} \rangle + \mathcal{O}\left( \frac{\delta}{\omega} \right) = \omega + \la {\bm Z}(\theta),\ {\bm p} \ra + \mathcal{O}\left(  \frac{\delta}{\omega} \right),
\label{A13}
\end{align}
where $\delta / \omega$ characterizes the order of the approximation error. We assume that $\delta / \omega$ is sufficiently small.

Next, taking an inner product of Eq.~(\ref{A9}) with ${\bm V}_k(\theta)$ with $k \geq 1$, we obtain
\begin{align}
\frac{d\theta}{dt} \left( - \frac{\lambda_k}{\omega} r_k + \frac{1}{\omega} \sum_j r_j \langle {\bm V}_k(\theta),\ {\bm J}(\theta) {\bm U}_j(\theta) \rangle \right) 
+
\frac{d r_k}{dt}
=
\sum_j r_j \langle {\bm V}_k(\theta),\ {\bm J}(\theta) {\bm U}_j(\theta) \rangle
+ 
\langle {\bm V}_k(\theta),\ {\bm p} \rangle
\end{align}
and, using Eq.~{(\ref{A12})},
\begin{align}
\frac{d r_k}{dt}
&= \lambda_k r_k + \langle {\bm V}_k(\theta),\ {\bm p} \rangle
+ \langle {\bm V}_0(\theta),\ {\bm p} \rangle \left( \frac{\lambda_k}{\omega} r_k - \frac{1}{\omega} \sum_j r_j \langle {\bm V}_k(\theta),\ {\bm J}(\theta) {\bm U}_j(\theta) \rangle \right) + {\mathcal{O}\left( \frac{ \delta^2}{\omega^2} \right)} .
\end{align}
Retaining only the terms up to $\mathcal{O}(\delta / \omega)$, we obtain
\begin{align}
\frac{d r_k}{dt}
&= \lambda_k r_k + \langle {\bm V}_k(\theta),\ {\bm p} \rangle + \mathcal{O}\left(  \frac{\delta}{\omega} \right) = \lambda_k r_k + \la {\bm I}_k(\theta),\ {\bm p} \ra + \mathcal{O}\left(  \frac{\delta}{\omega} \right).
\label{A16}
\end{align}
Equations~(\ref{A13}) and (\ref{A16}) give the first-order approximate phase-amplitude equations~(\ref{eq:pr_phi_r}) in the main text.

%%%%%%%%%%%%%%%%%
%%% Appendix A %%
%%%%%%%%%%%%%%%%%

\section{Tangential periodic input}

If we take the limit of large weights in the amplitude-penalty method, i.e., $k_i \to \infty$ for $i=1, ..., M$ in the objective function in Eq.~(\ref{eq:opt_pnt}), the optimal input waveform is expected to possess only the vector components tangent to the limit cycle. 
Here, we consider this limit and assume that the input waveform is always proportional to the tangent vector of the limit cycle, optimize its scalar coefficient, and compare the result with those of the amplitude-penalty method.

We assume an input waveform given in the form
\begin{align}
  \bm{q}(\Omega t) = \alpha(\Omega t){\bm{u}}_0( (\phi^*+\Omega t) / \omega ),
\end{align}
where $\alpha(\Omega t)$ is a temporally periodic scalar coefficient, and consider the following optimization problem for $\alpha(\Omega t)$:
\begin{align}
  \label{eq:opt_lins_tanonly}
  &\max_{\alpha}\ -\Gamma'(\phi^*),
  \cr
  &\mbox{s.t.}
  \quad
  \Delta + \Gamma(\phi^*) = 0,
  \quad
  \big[ \| \alpha(\Omega  t){\bm{u}}_0( (\phi^*+\Omega t) / \omega ) \|^2  \big]_t = P,
\end{align}
where the first constraint is for the phase-locking point and the second constraint is for the input power.
We introduce Lagrange multipliers $\mu$ and $\nu$ and consider a Lagrange function
\begin{align}
  C(\alpha, \mu, \nu) = -\Gamma'(\phi^*) + \mu \{\Delta + \Gamma(\phi^*)\} + \nu (P -  [ \| \alpha(\Omega  t){\bm{u}}_0((\phi^*+\Omega t)/\omega) \|^2  ]_t ).
\end{align}
From the extremum condition, we obtain
\begin{align}
  &
  \frac{{\delta}{C}}{{\delta}{\alpha}} = -\frac{1}{T_e}  \left\{ \la \bm{Z'}(\phi^* + \Omega t),   {\bm{u}}_0( (\phi^*  + \Omega t) /\omega )  \ra
  + \mu - 2 \nu \alpha(\Omega t) \| {\bm{u}}_0( (\Omega t + \phi^*)/\omega ) \|^2 \right\} = 0, \label{tan_condition1}
  \\
  &
  \frac{\partial{C}}{\partial{\mu}} = \Delta + [ \alpha(\Omega t) ]_t  = 0, \label{tan_condition2}
  \\
  &
  \frac{\partial{C}}{\partial{\nu}} = P -  [ \| \alpha(\Omega t){\bm{u}}_0( (\phi^*+\Omega t)/\omega ) \|^2]_t = 0, 
  \label{tan_condition3}
\end{align}
from which the optimal $\alpha(\Omega t)$ is obtained as
\begin{align}
  \alpha (\Omega  t) &= \frac{ -\la \bm{Z'}( \phi^* + \Omega t), {\bm{u}}_0( (\phi^* + \Omega t) /\omega ) \ra + \mu}{2\nu \| {\bm{u}}_0\|^2}, 
  \label{eq:tan_input} 
  \\ \cr
  \mu &= 
  \left( \left[ \frac{ \la \bm{Z}',  {\bm{u}}_0 \ra }{ \| {\bm{u}}_0 \|^2} \right]_t -2\nu\Delta \right) / \left[ \frac{1}{ \|{\bm{u}}_0 \|^2} \right]_t ,
  \\ \cr
  \nu &= \sqrt{\frac{- \left[ \frac{ \la \bm{Z}',  {\bm{u}}_0 \ra }{ \| {\bm{u}}_0 \|^2} \right]_t^2 + \left[ \frac{ \la \bm{Z}',  {\bm{u}}_0 \ra^2 }{ \| {\bm{u}}_0 \|^2} \right]_t \left[ \frac{1}{ \|{\bm{u}}_0 \|^2} \right]_t  }{4 \left( P\left[ \frac{1}{ \|{\bm{u}}_0 \|^2} \right]_t -\Delta^2 \right)}}. 
  \label{eq:tangent_lambda}
\end{align}
Here, for the existence of the optimal $\alpha(\Omega t)$, the argument inside the square root in Eq.~(\ref{eq:tangent_lambda}) should be positive, which restricts the allowed ranges of $P$ and $\Delta$.

\begin{figure} [!t]
  \begin{center}
    \includegraphics[width=0.85\hsize,clip]{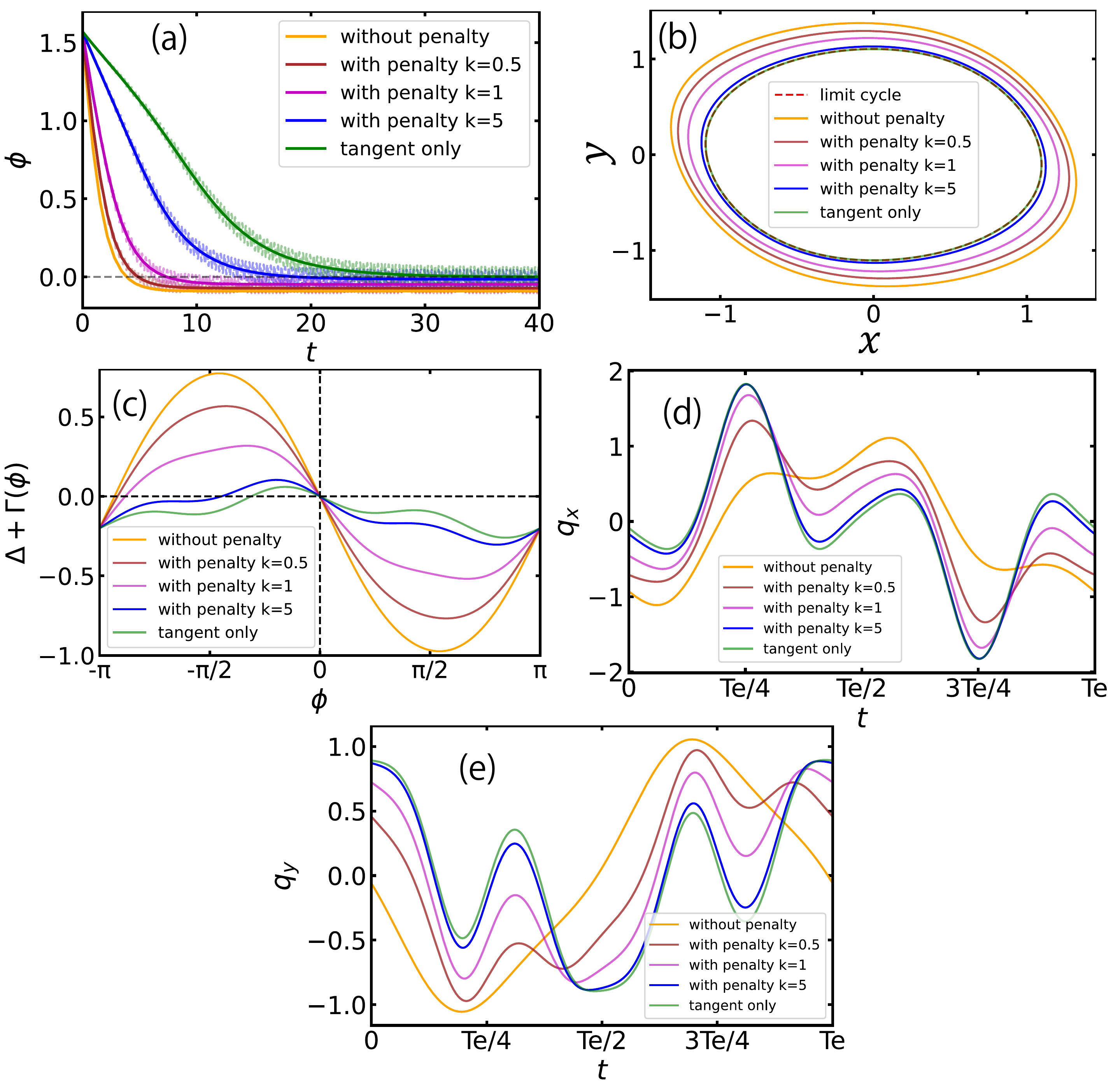}
    \caption{
      Optimal entrainment by the tangential input, compared with the amplitude-penalty method for the van der Pol model.
      (a) Evolution of the phase differences, (b) Trajectories on the phase plane, 
      (c) $\Delta + \Gamma(\phi)$ in Eq.~(\ref{phiT}), 
	  (d) Optimal inputs ($x$ component). 
      (e) Optimal inputs ($y$ component). 
      In (a), black dashed line shows the target phase-locking point $\phi^*=0$.
      In (b), red dashed curve shows the unperturbed limit cycle at $P=0$
    }
    \label{fig_8}
  \end{center}
\end{figure}

As an example, we consider the van der Pol model used in Sec.~V.
The input power and frequency are $P=1.0$ and $\Omega = \omega+0.1  \approx 10.04$, respectively.
The target phase-locking point is $\phi^*=0$ and the initial state of the system is chosen so that the initial phase difference is $\phi(0) = \pi / 2$.

Figure~\ref{fig_8} shows the results, where the entrainment by the  tangential input is compared with the entrainment by the amplitude-penalty method for several values of the weight $k_1 = k$.
The result for the amplitude-penalty method approaches the result for the tangential input as $k$ becomes larger.
For relatively small $k$, including $k = 0$ corresponding to the case without penalty (simple), the phase difference does not converge to the correct target value due to the amplitude deviations as shown in Fig.~\ref{fig_8}(b).
In contrast, for the case with a large amplitude penalty ($k=5$) and for the tangential case, the amplitude deviations are suppressed or do not arise and the phase difference converges to the correct value, $\phi^*=0$.

It is notable that the convergence in the tangential case is slower than in the case with the amplitude penalty.
This is because the optimal direction of the input waveform for stable entrainment generally possesses vector components in the amplitude directions, though of course too strong amplitude excitations lead to the breakdown of phase-only approximation. Thus, restricting the directions of the input waveform completely in the tangential direction hampers the realization of larger stability and leads to slower convergence. Allowing small but not too-large amplitude excitations is thus helpful for fast entrainment.

%%%%%%%%%%%%%%%%%
%%% Appendix B %%
%%%%%%%%%%%%%%%%%

\section{Numerical methods for Floquet exponents and eigenvectors}

In this section, we explain the details of the numerical method used to calculate the Floquet eigenvalues and eigenvectors, including the case with complex eigenvalues.
The eigenvalues and eigenvectors are calculated in the increasing order of $i$ from $i=0$.

%%%%%%%%%%%%%%%%%%%%%%%%%%%%%%%%%%%%%%%%%%%%%%%%
\subsection{Eigenvector for $i=0$}
%%%%%%%%%%%%%%%%%%%%%%%%%%%%%%%%%%%%%%%%%%%%%%%%

The zeroth eigenvalue is $\lambda_0 = 0$. The associated right eigenvector ${\bm{u}}_0(t)$ can be taken as the tangent vector along the limit cycle~\cite{kuramoto1984chemical}, i.e.,
\begin{align}
  {\bm{u}}_0(t) = \frac{1}{\omega}\frac{d{\bm X}_0(t)}{dt} = \frac{1}{\omega}\bm{F}({\bm X}_0(t)),
\end{align}
where the length scale of ${\bm u}_0(t)$ is chosen to be consistent with the convention of the phase reduction theory.
The zeroth left eigenvector ${\bm{v}}_0(t)$, which is equivalent to the PSF, can be calculated by the adjoint method~\cite{brown2004phase, nakao2016phase}, namely, by backward integration of the adjoint linear equation
\begin{align}
  \label{eq:v_0}
  \frac{d\bm{y}(t)}{dt} = -{\bm J}({\bm X}_0(t))^\dag\bm{y}(t)
\end{align}
from an arbitrary final condition.
Since $ \lambda_0 = 0 > \mbox{Re}~{\lambda_1} \geq \ldots  \geq  \mbox{Re}~{\lambda_{N-1}} $, $\bm{y}(t)$ eventually converges to a periodic solution satisfying $\bm{y}(t) \propto  {\bm{v}}_0(t)$ after a transient.
Since ${\bm{v}}_0(t)$ should satisfy the normalization condition $\la {\bm{v}}_0(t),  {\bm{u}}_0(t) \ra = 1$, we normalize $\bm{y}(t)$ at intervals of $T$ as
\begin{align}
  \bm{y}(t) \rightarrow \frac{\bm{y}(t)}{\langle \bm{y}(t), {\bm{u}}_0(t) \rangle}
\end{align}
during the calculation.

%%%%%%%%%%%%%%%%%%%%%%%%%%%%%%%%%%%%%%%%%%%%%%%%
\subsection{Eigenvalues and eigenvectors for $i \geq 1$ (real)}
%%%%%%%%%%%%%%%%%%%%%%%%%%%%%%%%%%%%%%%%%%%%%%%%

We assume that the eigenvalue and the eigenvectors $\{ \lambda_i, {\bm{u}}_i(t), {\bm{v}}_i(t)\}$ ($i \geq 1$) are real and simple, and the eigenvalues and eigenvectors for $j = 0, 1, ..., i-1$ are already obtained.
We numerically integrate the linearized equation
\begin{align}
  \frac{d}{dt}\bm{y}(t) = {\bm J}({\bm X}_0(t)) \bm{y}(t)
\end{align}
from an arbitrary initial condition. During the calculation, we remove unnecessary vector components for $j=0, 1, ..., i-1$ to obtain the correct $i$th vector components (the vector components for $j=i+1, ..., N-1$ have larger decay rates and automatically vanish).
Namely, we subtract ${\bm{u}}_j(t)$ ($j= 0,1, \ldots, i-1$) at each time step as
\begin{align}
  \label{eq:Floquet_eigenvectors(Real)_u1}
  \bm{y}(t) \rightarrow \bm{y}(t) -  \sum_{j = 0}^{i-1} \la  {\bm{v}}_{j}(t),  \bm{y}(t) \ra {\bm{u}}_j(t).
\end{align}
Since $ \mbox{Re}~{\lambda_i} > \mbox{Re}~{\lambda_{i+1}}  \geq  \ldots  \geq  \mbox{Re}~{\lambda_{N-1}} $, the solution converges to $\bm{y}(t) \propto  {\bm{u}}_i(t)$ after a transient.
We assume the length of $ {\bm{u}}_i(t)$ to be $| {\bm u}_i(t) | = 1$ (this can be arbitrary chosen and determines the scale of the $i$th amplitude $r_i$) and normalize ${\bm y}(t)$ to satisfy 
$ {\la \bm{y}(t), \bm{y}(t) \ra}  = 1$ at each time step.
After the transient, we take ${\bm y}(t)$ as a new initial state ${\bm y}(0)$ and calculate the eigenvalue $\lambda_i$ as the growth rate of ${\bm y}$ during one period of oscillation as
\begin{align}
  \lambda_i = \frac{1}{T}\ln \frac{\bm{y}(T)}{\bm{y}(0)}.
\end{align}
The right eigenvector ${\bm{u}}_i(t)$ can then be calculated by integrating
\begin{align}
  \frac{d}{dt}\bm{y}(t) = [{\bm J}({\bm X}_0(t)) - \lambda_i] \bm{y}(t)
\end{align}
from ${\bm y}(0)$ for one period, which gives ${\bm{u}}_i(t) = \bm{y}(t)~(0 \leq t < T)$.

We next calculate the left eigenvector ${\bm{v}}_i(t)$.
We integrate the following equation from an arbitrary final condition backward in time:
\begin{align}
  \frac{d}{dt}\bm{y}(t) = -{\bm J}({\bm X}_0(t))^\dag\bm{y}(t). 
  \label{eq:v_1[1]}
\end{align}
During the calculation, we subtract the vector components in the directions ${\bm{v}}_j(t)$ ($j= 0,1, \ldots, j-1$) at each time step as
\begin{align}
  \label{eq:v_1[2]}
  \bm{y}(t) \rightarrow \bm{y}(t)  -  \sum_{j = 0}^{i-1} \la  \bm{y}(t), {\bm{u}}_{j}(t) \ra {\bm{v}}_j(t).
\end{align}
Since $ \mbox{Re}~{\lambda_i} > \mbox{Re}~{\lambda_{i+1}}  \geq  \ldots  \geq  \mbox{Re}~{\lambda_{N-1}}$, we eventually obtain $\bm{y}(t) \propto  {\bm{v}}_i(t)$ after the transient.
In order to satisfy the normalization condition
$\la {\bm{v}}_i(t),  {\bm{u}}_i(t) \ra = 1$,
we periodically normalize $\bm{y}(t)$ at intervals of $T$ as follows:
\begin{align}
  \bm{y}(t) \rightarrow \frac{\bm{y}(t)}{\langle \bm{y}(t), {\bm{u}}_i(t) \rangle}.
\end{align}
By using ${\bm y}(t)$ sufficiently after the initial transient as a new final state ${\bm y}(T)$, the left eigenvector can be calculated by backward integration of
\begin{align}
  \label{eq:adj_vi}
  \frac{d\bm{y}(t)}{dt} = -[{\bm J}({\bm X}_0(t))^\dag - \lambda^*_i ]\bm{y}(t)
\end{align}
for one period $(0 \leq t < T)$ as ${\bm{v}}_i(t) = \bm{y}(t)$.

%%%%%%%%%%%%%%%%%%%%%%%%%%%%%%%%%%%%%%%%%%%%%%%%
\subsection{Eigenvalues and eigenvectors for $i \geq 1$ (complex)}
%%%%%%%%%%%%%%%%%%%%%%%%%%%%%%%%%%%%%%%%%%%%%%%%

We assume that $\{\lambda_i, {\bm{u}}_i, {\bm{v}}_i \}$ and $\{\lambda_{i+1}, {\bm{u}}_{i+1}, {\bm{v}}_{i+1} \}$ for some $i \geq 1$ are mutually complex conjugate,
namely, $\lambda_{i +1} =  \lambda^*_{i}$, ${\bm{v}}_{i+1} = {\bm{v}}^*_{i}$, and ${\bm{u}}_{i+1} = {\bm{u}}^*_{i}$, and also $\mbox{Re}\ \lambda_{i}, \lambda_{i+1} > \mbox{Re}\ \lambda_{i+2}$.
We also assume that the eigenvalues and eigenvectors for $j = 0, 1, ..., i-1$ are already obtained.
In this case, we cannot use the numerical method for the real vectors, so we calculate the complex eigenvalues and eigenvectors directly from the monodromy matrix $\bm{M} = \exp( \bm{\Lambda} T)$, denoted as $\bm{M} = [\bm{m}_0, \bm{m}_1, \cdots, \bm{m}_{N-1}]$ where $\{ {\bm m}_k \in {\mathbb R}^{N} \}_{k=0, ..., N-1}$ are column vectors. By solving 
\begin{align}
  \label{eq:jacob}
  \frac{d}{dt}\bm{y}(t) = {\bm J}({\bm X}_0(t)) \bm{y}(t)
\end{align}
from an arbitrary initial condition $\bm{y}(0)$ at $t=0$ for one oscillation period $T$, we obtain $\bm{y}(T) =  \exp( \bm{\Lambda} T) \bm{y}(0)$.
By choosing a unit vector in the $k$th direction $(k=0, \cdots N-1)$ as the initial condition, i.e., $\bm{y}(0)=\bm{e}_k$, we can obtain $\bm{y}(T)=\bm{m}_i~~(i=0, \cdots N-1)$. 
From the numerically evaluated $\bm{M}$ and $\bm{M^\dag}$, we can evaluate 
$\{ \lambda_i, {\bm{u}}_i, {\bm{v}}_i \}$ and $\{ \lambda_{i+1}, {\bm{u}}_{i+1}, {\bm{v}}_{i+1} \}$, i.e., the Floquet exponents and the Floquet eigenvectors at ${\bm X}_0(0)$.

The right eigenvectors ${\bm{u}}_{i'}(t)$ $(i' = i, i + 1)$ for $0 \leq t < T$ can be obtained by integrating
\begin{align}
  \label{eq:u_ceig}
  \frac{d}{dt} \bm{y}(t) = [{\bm J}({\bm X}_0(t)) - \lambda_{i'}] \bm{y}(t)
\end{align}
from the initial condition $\bm{y}(0) = {\bm{u}}_{i'}$ $(i' = i, i + 1)$ for one oscillation period $T$. During the calculation, we subtract the components in the directions ${\bm{u}}_j(t)$ $(j= 0,1, \ldots, i-1)$ at each time step as
\begin{align}
  \bm{y}(t) \rightarrow \bm{y}(t) -  \sum_{j = 0}^{i-1} \la  {\bm{v}}_{j}(t),  \bm{y}(t) \ra {\bm{u}}_j(t).
\end{align}
Next, the left eigenvectors ${\bm{v}}_{i'}(t)$ $(i' = i, i + 1)$ for $0 \leq t < T$ can be calculated by integrating
\begin{align}
  \label{eq:v_ceig}
  \frac{d}{dt}  \bm{y}(t) = -[{\bm J}({\bm X}_0(t))^\dag - \lambda_{i'}^*] \bm{y}(t).
\end{align}
backward from the final condition $\bm{y}(T) = {\bm{v}}_{i'}$ $(i' = i, i + 1)$. During the calculation, we subtract the coefficients in the directions of ${\bm{v}}_j(t)$ $(j= 0,1, \ldots, i-1)$ at each time step by
\begin{align}
  \label{eq:v_elim}
  \bm{y}(t) \rightarrow \bm{y}(t)  -  \sum_{j = 0}^{i-1} \la  \bm{y}(t), {\bm{u}}_{j}(t) \ra {\bm{v}}_j(t).
\end{align}

In this method, we need to obtain an accurate monodromy matrix ${\bm M}$ by using sufficiently small time steps for numerical integration. Also, it is not guaranteed in this method that the vector components in the directions of ${\bm{u}}_{i}(t)$ and ${\bm{u}}_{i+1}(t)$
and those in the directions of ${\bm{v}}_{i}(t)$ and ${\bm{v}}_{i+1}(t)$ do not mix due to numerical errors. The validity of the numerical results can be confirmed by checking the bi-orthonormality relation $ \la \bm{v}_{j}(t), \bm{u}_{k}(t) \ra = {\delta}_{jk}$ for $j, k = i, i+1$ after the calculation.
If the numerical errors are non-negligible, we may need to introduce an additional bi-orthonormalization procedure to separate the vector components from each other.
For the two examples used in this study, these numerical errors were negligible in our calculations. 
We also note that the method in this subsection can also be used to calculate real eigenvalues and eigenvectors, but the method in the previous subsection is numerically more accurate.


\begin{thebibliography}{10}
	
	\bibitem{pantaleone2002synchronization}
	James Pantaleone.
	\newblock Synchronization of metronomes.
	\newblock {\em American Journal of Physics}, 70(10):992--1000, 2002.
	
	\bibitem{winfree1972spiral}
	Arthur~T Winfree.
	\newblock Spiral waves of chemical activity.
	\newblock {\em Science}, 175(4022):634--636, 1972.
	
	\bibitem{buck1968mechanism}
	John Buck and Elisabeth Buck.
	\newblock Mechanism of rhythmic synchronous flashing of fireflies: Fireflies of
	southeast asia may use anticipatory time-measuring in synchronizing their
	flashing.
	\newblock {\em Science}, 159(3821):1319--1327, 1968.
	
	\bibitem{buck1976synchronous}
	John Buck and Elisabeth Buck.
	\newblock Synchronous fireflies.
	\newblock {\em Scientific American}, 234(5):74--85, 1976.
	
	\bibitem{ermentrout1984beyond}
	G~Bard Ermentrout and John Rinzel.
	\newblock Beyond a pacemaker's entrainment limit: phase walk-through.
	\newblock {\em American Journal of Physiology-Regulatory, Integrative and
		Comparative Physiology}, 246(1):R102--R106, 1984.
	
	\bibitem{goldbeter1995model}
	Albert Goldbeter.
	\newblock A model for circadian oscillations in the drosophila period protein
	(per).
	\newblock {\em Proceedings of the Royal Society of London. Series B: Biological
		Sciences}, 261(1362):319--324, 1995.
	
	\bibitem{leloup1999limit}
	Jean-Christophe Leloup, Didier Gonze, and Albert Goldbeter.
	\newblock Limit cycle models for circadian rhythms based on transcriptional
	regulation in drosophila and neurospora.
	\newblock {\em Journal of biological rhythms}, 14(6):433--448, 1999.
	
	\bibitem{winfree2001geometry}
	Arthur~T Winfree.
	\newblock {\em The geometry of biological time}.
	\newblock Springer, New York, 2001.
	
	\bibitem{kuramoto1984chemical}
	Yoshiki {K}uramoto.
	\newblock {\em Chemical oscillations, waves, and turbulence}.
	\newblock Springer, Berlin, 1984.
	
	\bibitem{ermentrout2010mathematical}
	G~Bard Ermentrout and David~H Terman.
	\newblock {\em Mathematical foundations of neuroscience}.
	\newblock Springer, New York, 2010.
	
	\bibitem{pikovsky2001synchronization}
	Arkady Pikovsky, Michael Roblum, and J{\"u}rgen Kurths.
	\newblock {\em Synchronization: a universal concept in nonlinear sciences}.
	\newblock Cambridge University Press, Cambridge, 2001.
	
	\bibitem{glass1988clocks}
	Leon Glass and Michael~C Mackey.
	\newblock {\em From clocks to chaos: the rhythms of life}.
	\newblock Princeton University Press, Princeton, 1988.
	
	\bibitem{strogatz1994nonlinear}
	Steven~H Strogatz.
	\newblock {\em Nonlinear dynamics and chaos}.
	\newblock Westview Press, 1994.
	
	\bibitem{kawasaki2010millimeter}
	Kenichi Kawasaki, Yoshiyuki Akiyama, Kenji Komori, Masahiro Uno, Hidenori
	Takeuchi, Tomoari Itagaki, Yasufumi Hino, Yoshinobu Kawasaki, Katsuhisa Ito,
	and Ali Hajimiri.
	\newblock A millimeter-wave intra-connect solution.
	\newblock {\em IEEE Journal of Solid-State Circuits}, 45(12):2655--2666, 2010.
	
	\bibitem{daryoush1990optical}
	Afshin~S Daryoush.
	\newblock Optical synchronization of millimeter-wave oscillators for
	distributed architecture.
	\newblock {\em IEEE Transactions on microwave theory and techniques},
	38(5):467--476, 1990.
	
	\bibitem{nagashima2014locking}
	Tomoharu Nagashima, Xiuqin Wei, Hisa-Aki Tanaka, and Hiroo Sekiya.
	\newblock Locking range derivations for injection-locked class-e oscillator
	applying phase reduction theory.
	\newblock {\em IEEE Transactions on Circuits and Systems I: Regular Papers},
	61(10):2904--2911, 2014.
	
	\bibitem{wilson2017spatiotemporal}
	Dan Wilson and Jeff Moehlis.
	\newblock Spatiotemporal control to eliminate cardiac alternans using isostable
	reduction.
	\newblock {\em Physica D: Nonlinear Phenomena}, 342:32--44, 2017.
	
	\bibitem{monga2019optimal}
	Bharat Monga and Jeff Moehlis.
	\newblock Optimal phase control of biological oscillators using augmented phase
	reduction.
	\newblock {\em Biological Cybernetics}, 113(1-2):161--178, 2019.
	
	\bibitem{stone2019application}
	Julia~E Stone, Xavier~L Aubert, Henning Maass, Andrew~JK Phillips, Michelle
	Magee, Mark~E Howard, Steven~W Lockley, Shantha~MW Rajaratnam, and Tracey~L
	Sletten.
	\newblock Application of a limit-cycle oscillator model for prediction of
	circadian phase in rotating night shift workers.
	\newblock {\em Scientific Reports}, 9(1):1--12, 2019.
	
	\bibitem{nakao2016phase}
	Hiroya Nakao.
	\newblock Phase reduction approach to synchronisation of nonlinear oscillators.
	\newblock {\em Contemporary Physics}, 57(2):188--214, 2016.
	
	\bibitem{monga2019phase}
	Bharat Monga, Dan Wilson, Tim Matchen, and Jeff Moehlis.
	\newblock Phase reduction and phase-based optimal control for biological
	systems: a tutorial.
	\newblock {\em Biological Cybernetics}, 113(1-2):11--46, 2019.
	
	\bibitem{kuramoto2019concept}
	Yoshiki {K}uramoto and Hiroya Nakao.
	\newblock On the concept of dynamical reduction: the case of coupled
	oscillators.
	\newblock {\em Philosophical Transactions of the Royal Society A},
	377(2160):20190041, 2019.
	
	\bibitem{acebron2005kuramoto}
	Juan~A Acebr{\'o}n, Luis~L Bonilla, Conrad J~P{\'e}rez Vicente, F{\'e}lix
	Ritort, and Renato Spigler.
	\newblock The {K}uramoto model: A simple paradigm for synchronization
	phenomena.
	\newblock {\em Reviews of Modern Physics}, 77(1):137, 2005.
	
	\bibitem{strogatz2000kuramoto}
	Steven~H Strogatz.
	\newblock From {K}uramoto to crawford: exploring the onset of synchronization
	in populations of coupled oscillators.
	\newblock {\em Physica D: Nonlinear Phenomena}, 143(1-4):1--20, 2000.
	
	\bibitem{shirasaka2017phase2}
	Sho Shirasaka, Wataru Kurebayashi, and Hiroya Nakao.
	\newblock Phase reduction theory for hybrid nonlinear oscillators.
	\newblock {\em Physical Review E}, 95(1):012212, 2017.
	
	\bibitem{kotani2012adjoint}
	Kiyoshi Kotani, Ikuhiro Yamaguchi, Yutaro Ogawa, Yasuhiko Jimbo, Hiroya Nakao,
	and G~Bard Ermentrout.
	\newblock Adjoint method provides phase response functions for delay-induced
	oscillations.
	\newblock {\em Physical Review Letters}, 109(4):044101, 2012.
	
	\bibitem{kawamura2013collective}
	Yoji Kawamura and Hiroya Nakao.
	\newblock Collective phase description of oscillatory convection.
	\newblock {\em Chaos: An Interdisciplinary Journal of Nonlinear Science},
	23(4):043129, 2013.
	
	\bibitem{nakao2014phase}
	Hiroya Nakao, Tatsuo Yanagita, and Yoji Kawamura.
	\newblock Phase-reduction approach to synchronization of spatiotemporal rhythms
	in reaction-diffusion systems.
	\newblock {\em Physical Review X}, 4(2):021032, 2014.
	
	\bibitem{kato2019semiclassical}
	Yuzuru Kato, Naoki Yamamoto, and Hiroya Nakao.
	\newblock Semiclassical phase reduction theory for quantum synchronization.
	\newblock {\em Phys. Rev. Research}, 1:033012, Oct 2019.
	
	\bibitem{moehlis2006optimal}
	Jeff Moehlis, Eric Shea-Brown, and Herschel Rabitz.
	\newblock Optimal inputs for phase models of spiking neurons.
	\newblock {\em Journal of Computational and Nonlinear Dynamics}, 1(4):358--367,
	2006.
	
	\bibitem{dasanayake2011optimal}
	Isuru Dasanayake and Jr-Shin Li.
	\newblock Optimal design of minimum-power stimuli for phase models of neuron
	oscillators.
	\newblock {\em Physical Review E}, 83(6):061916, 2011.
	
	\bibitem{zlotnik2012optimal}
	Anatoly Zlotnik and Jr-Shin Li.
	\newblock Optimal entrainment of neural oscillator ensembles.
	\newblock {\em Journal of Neural Engineering}, 9(4):046015, 2012.
	
	\bibitem{li2013control}
	Jr-Shin Li, Isuru Dasanayake, and Justin Ruths.
	\newblock Control and synchronization of neuron ensembles.
	\newblock {\em IEEE Transactions on automatic control}, 58(8):1919--1930, 2013.
	
	\bibitem{harada2010optimal}
	Takahiro Harada, Hisa-Aki Tanaka, Michael~J Hankins, and Istv{\'a}n~Z Kiss.
	\newblock Optimal waveform for the entrainment of a weakly forced oscillator.
	\newblock {\em Physical Review Letters}, 105(8):088301, 2010.
	
	\bibitem{tanaka2014optimal}
	Hisa-Aki Tanaka.
	\newblock Optimal entrainment with smooth, pulse, and square signals in weakly
	forced nonlinear oscillators.
	\newblock {\em Physica D: Nonlinear Phenomena}, 288:1--22, 2014.
	
	\bibitem{tanaka2015optimal}
	Hisa-Aki Tanaka, Isao Nishikawa, J{\"u}rgen Kurths, Yifei Chen, and
	Istv{\'a}n~Z Kiss.
	\newblock Optimal synchronization of oscillatory chemical reactions with
	complex pulse, square, and smooth waveforms signals maximizes tsallis
	entropy.
	\newblock {\em EPL (Europhysics Letters)}, 111(5):50007, 2015.
	
	\bibitem{zlotnik2013optimal}
	Anatoly Zlotnik, Yifei Chen, Istv{\'a}n~Z Kiss, Hisa-Aki Tanaka, and Jr-Shin
	Li.
	\newblock Optimal waveform for fast entrainment of weakly forced nonlinear
	oscillators.
	\newblock {\em Physical Review Letters}, 111(2):024102, 2013.
	
	\bibitem{shirasaka2017optimizing}
	Sho Shirasaka, Nobuhiro Watanabe, Yoji Kawamura, and Hiroya Nakao.
	\newblock Optimizing stability of mutual synchronization between a pair of
	limit-cycle oscillators with weak cross coupling.
	\newblock {\em Physical Review E}, 96(1):012223, 2017.
	
	\bibitem{watanabe2019optimization}
	Nobuhiro Watanabe, Yuzuru Kato, Sho Shirasaka, and Hiroya Nakao.
	\newblock Optimization of linear and nonlinear interaction schemes for stable
	synchronization of weakly coupled limit-cycle oscillators.
	\newblock {\em Physical Review E}, 100:042205, Oct 2019.
	
	\bibitem{pikovsky2015maximizing}
	Arkady Pikovsky.
	\newblock Maximizing coherence of oscillations by external locking.
	\newblock {\em Physical Review Letters}, 115(7):070602, 2015.
	
	\bibitem{zlotnik2016phase}
	Anatoly Zlotnik, Raphael Nagao, Istv{\'a}n~Z Kiss, and Jr-Shin Li.
	\newblock Phase-selective entrainment of nonlinear oscillator ensembles.
	\newblock {\em Nature Communications}, 7:10788, 2016.
	
	\bibitem{monga2018synchronizing}
	Bharat Monga, Gary Froyland, and Jeff Moehlis.
	\newblock Synchronizing and desynchronizing neural populations through phase
	distribution control.
	\newblock In {\em 2018 Annual American Control Conference (ACC)}, pages
	2808--2813. IEEE, 2018.
	
	\bibitem{kuritz2019ensemble}
	Karsten Kuritz, Shen Zeng, and Frank Allg{\"o}wer.
	\newblock Ensemble controllability of cellular oscillators.
	\newblock {\em IEEE Control Systems Letters}, 3(2):296--301, 2019.
	
	\bibitem{monga2019phase2}
	Bharat Monga and Jeff Moehlis.
	\newblock Phase distribution control of a population of oscillators.
	\newblock {\em Physica D: Nonlinear Phenomena}, 398:115--129, 2019.
	
	\bibitem{kato2021optimization}
	Yuzuru Kato, Anatoly Zlotnik, Jr-Shin Li, and Hiroya Nakao.
	\newblock Optimization of periodic input waveforms for global entrainment of
	weakly forced limit-cycle oscillators.
	\newblock {\em Nonlinear Dynamics, in press (arXiv preprint arXiv:2103.02880)},
	2021.
	
	\bibitem{kawamura2017optimizing}
	Yoji Kawamura, Sho Shirasaka, Tatsuo Yanagita, and Hiroya Nakao.
	\newblock Optimizing mutual synchronization of rhythmic spatiotemporal patterns
	in reaction-diffusion systems.
	\newblock {\em Physical Review E}, 96(1):012224, 2017.
	
	\bibitem{yamaguchi2021network}
	Hiroya Nakao, Katsunori Yamaguchi, Shingo Katayama, and Tatsuo Yanagita.
	\newblock Sparse optimization of mutual synchronization in collectively
	oscillating networks.
	\newblock {\em Chaos: An Interdisciplinary Journal of Nonlinear Science},
	31(6):063113, 2021.
	
	\bibitem{kato2020semiclassical}
	Yuzuru Kato and Hiroya Nakao.
	\newblock Semiclassical optimization of entrainment stability and phase
	coherence in weakly forced quantum limit-cycle oscillators.
	\newblock {\em Physical Review E}, 101(1):012210, 2020.
	
	\bibitem{mezic2005spectral}
	Igor Mezi{\'c}.
	\newblock Spectral properties of dynamical systems, model reduction and
	decompositions.
	\newblock {\em Nonlinear Dynamics}, 41(1-3):309--325, 2005.
	
	\bibitem{mezic2013analysis}
	Igor Mezi{\'c}.
	\newblock Analysis of fluid flows via spectral properties of the {K}oopman
	operator.
	\newblock {\em Annual Review of Fluid Mechanics}, 45:357--378, 2013.
	
	\bibitem{mauroy2013isostables}
	Alexandre Mauroy, Igor Mezi{\'c}, and Jeff Moehlis.
	\newblock Isostables, isochrons, and {K}oopman spectrum for the action--angle
	representation of stable fixed point dynamics.
	\newblock {\em Physica D: Nonlinear Phenomena}, 261:19--30, 2013.
	
	\bibitem{mauroy2014global}
	Alexandre Mauroy, Blane Rhoads, Jeff Moehlis, and Igor Mezic.
	\newblock Global isochrons and phase sensitivity of bursting neurons.
	\newblock {\em SIAM Journal on Applied Dynamical Systems}, 13(1):306--338,
	2014.
	
	\bibitem{wilson2016isostable}
	Dan Wilson and Jeff Moehlis.
	\newblock Isostable reduction of periodic orbits.
	\newblock {\em Physical Review E}, 94(5):052213, 2016.
	
	\bibitem{mauroy2016global}
	Alexandre Mauroy and Igor Mezi{\'c}.
	\newblock Global stability analysis using the eigenfunctions of the {K}oopman
	operator.
	\newblock {\em IEEE Transactions on Automatic Control}, 61(11):3356--3369,
	2016.
	
	\bibitem{mauroy2018global}
	Alexandre Mauroy and Igor Mezi{\'c}.
	\newblock Global computation of phase-amplitude reduction for limit-cycle
	dynamics.
	\newblock {\em Chaos: An Interdisciplinary Journal of Nonlinear Science},
	28(7):073108, 2018.
	
	\bibitem{shirasaka2017phase}
	Sho Shirasaka, Wataru Kurebayashi, and Hiroya Nakao.
	\newblock Phase-amplitude reduction of transient dynamics far from attractors
	for limit-cycling systems.
	\newblock {\em Chaos: An Interdisciplinary Journal of Nonlinear Science},
	27(2):023119, 2017.
	
	\bibitem{shirasaka2020phase}
	Sho Shirasaka, Wataru Kurebayashi, and Hiroya Nakao.
	\newblock Phase-amplitude reduction of limit cycling systems.
	\newblock In {\em The {K}oopman Operator in Systems and Control}, pages
	383--417. Springer, 2020.
	
	\bibitem{kotani2020nonlinear}
	Kiyoshi Kotani, Yutaro Ogawa, Sho Shirasaka, Akihiko Akao, Yasuhiko Jimbo, and
	Hiroya Nakao.
	\newblock Nonlinear phase-amplitude reduction of delay-induced oscillations.
	\newblock {\em Physical Review Research}, 2(3):033106, 2020.
	
	\bibitem{nakao2021phaseamplitude}
	Hiroya Nakao.
	\newblock Phase and amplitude description of complex oscillatory patterns in
	reaction-diffusion systems.
	\newblock In {\em Physics of Biological Oscillators}, pages 11--27. Springer,
	2021.
	
	\bibitem{wilson2018greater}
	Dan Wilson and Bard Ermentrout.
	\newblock Greater accuracy and broadened applicability of phase reduction using
	isostable coordinates.
	\newblock {\em Journal of Mathematical Biology}, 76(1):37--66, 2018.
	
	\bibitem{wilson2021optimal}
	Dan Wilson.
	\newblock Optimal control of oscillation timing and entrainment using large
	magnitude inputs: An adaptive phase-amplitude-coordinate-based approach.
	\newblock {\em arXiv preprint arXiv:2102.04535}, 2021.
	
	\bibitem{ermentrout1996type}
	Bard Ermentrout.
	\newblock Type i membranes, phase resetting curves, and synchrony.
	\newblock {\em Neural Computation}, 8(5):979--1001, 1996.
	
	\bibitem{brown2004phase}
	Eric Brown, Jeff Moehlis, and Philip Holmes.
	\newblock On the phase reduction and response dynamics of neural oscillator
	populations.
	\newblock {\em Neural Computation}, 16(4):673--715, 2004.
	
	\bibitem{jordan1999nonlinear}
	Dominic~William Jordan and Peter Smith.
	\newblock {\em Nonlinear ordinary differential equations: an introduction to
		dynamical systems}, volume~2.
	\newblock Oxford University Press, USA, 1999.
	
	\bibitem{guckenheimer2013nonlinear}
	John Guckenheimer and Philip Holmes.
	\newblock {\em Nonlinear oscillations, dynamical systems, and bifurcations of
		vector fields}, volume~42.
	\newblock Springer Science \& Business Media, 2013.
	
	\bibitem{hoppensteadt1997weakly}
	Frank~C Hoppensteadt and Eugene~M Izhikevich.
	\newblock {\em Weakly Connected Neural Networks}.
	\newblock Springer, 1997.
	
	\bibitem{van1927frequency}
	Balth Van~der Pol and Jan Van Der~Mark.
	\newblock Frequency demultiplication.
	\newblock {\em Nature}, 120(3019):363--364, 1927.
	
	\bibitem{van1927vii}
	Balth Van~{d}er Pol.
	\newblock Vii. forced oscillations in a circuit with non-linear
	resistance.(reception with reactive triode).
	\newblock {\em The London, Edinburgh, and Dublin Philosophical Magazine and
		Journal of Science}, 3(13):65--80, 1927.
	
	\bibitem{willamowski1980irregular}
	K-D Willamowski and OE~R{\"o}ssler.
	\newblock Irregular oscillations in a realistic abstract quadratic mass action
	system.
	\newblock {\em Zeitschrift f{\"u}r Naturforschung A}, 35(3):317--318, 1980.
	
	\bibitem{boland2009limit}
	Richard~P Boland, Tobias Galla, and Alan~J McKane.
	\newblock Limit cycles, complex floquet multipliers, and intrinsic noise.
	\newblock {\em Physical Review E}, 79(5):051131, 2009.
	
	\bibitem{geysermans1996particle}
	Pascale Geysermans and Florence Baras.
	\newblock Particle simulation of chemical chaos.
	\newblock {\em The Journal of Chemical Physics}, 105(4):1402--1408, 1996.
	
	\bibitem{aguda1988dynamic}
	Baltazar~D Aguda and Bruce~L Clarke.
	\newblock Dynamic elements of chaos in the willamowski--r{\"o}ssler network.
	\newblock {\em The Journal of Chemical Physics}, 89(12):7428--7434, 1988.
	
	\bibitem{kurebayashi2013phase}
	Wataru Kurebayashi, Sho Shirasaka, and Hiroya Nakao.
	\newblock Phase reduction method for strongly perturbed limit cycle
	oscillators.
	\newblock {\em Physical Review Letters}, 111(21):214101, 2013.
	
	\bibitem{cui2009functional}
	Jianxia Cui, Carmen~C Canavier, and Robert~J Butera.
	\newblock Functional phase response curves: a method for understanding
	synchronization of adapting neurons.
	\newblock {\em Journal of Neurophysiology}, 102(1):387--398, 2009.
	
	\bibitem{castejon2013phase}
	Oriol Castej{\'o}n, Antoni Guillamon, and Gemma Huguet.
	\newblock Phase-amplitude response functions for transient-state stimuli.
	\newblock {\em The Journal of Mathematical Neuroscience}, 3(1):1--26, 2013.
	
	\bibitem{rosenblum2019numerical}
	Michael Rosenblum and Arkady Pikovsky.
	\newblock Numerical phase reduction beyond the first order approximation.
	\newblock {\em Chaos: An Interdisciplinary Journal of Nonlinear Science},
	29(1):011105, 2019.
	
	\bibitem{wilson2015optimal}
	Dan Wilson, Abbey~B Holt, Theoden~I Netoff, and Jeff Moehlis.
	\newblock Optimal entrainment of heterogeneous noisy neurons.
	\newblock {\em Frontiers in Neuroscience}, 9:192, 2015.
	
	\bibitem{wilson2014energy}
	Dan Wilson and Jeff Moehlis.
	\newblock An energy-optimal approach for entrainment of uncertain circadian
	oscillators.
	\newblock {\em Biophysical Journal}, 107(7):1744--1755, 2014.
	
	\bibitem{monga2020supervised}
	Bharat Monga and Jeff Moehlis.
	\newblock Supervised learning algorithms for controlling underactuated
	dynamical systems.
	\newblock {\em Physica D: Nonlinear Phenomena}, 412:132621, 2020.
	
	\bibitem{wilson2020data}
	Dan Wilson.
	\newblock A data-driven phase and isostable reduced modeling framework for
	oscillatory dynamical systems.
	\newblock {\em Chaos: An Interdisciplinary Journal of Nonlinear Science},
	30(1):013121, 2020.
	
	\bibitem{wilson2021data}
	Dan Wilson.
	\newblock Data-driven inference of high-accuracy isostable-based dynamical
	models in response to external inputs.
	\newblock {\em Chaos: An Interdisciplinary Journal of Nonlinear Science},
	31:073103, 2021.
	
	\bibitem{github}
	Shohei Takata.
	\newblock {GitHub}: {S}hohei-{T}akata/{F}ast-optimal-entrainment.
	\newblock {\em https://github.com/Shohei-Takata/Fast-optimal-entrainment}.
	
\end{thebibliography}
\end{document}